\documentclass[pra,superscriptaddress,showpacs,papersize=a4paper,twocolumn,floatfix]{revtex4-1}

\usepackage[pdftex]{graphicx}
\usepackage{bm}
\usepackage{etoolbox}
\usepackage[usenames,dvipsnames]{color}
\usepackage{mathtools}
\usepackage{amsmath}
\usepackage{amsfonts}
\usepackage{amssymb}
\graphicspath{{}}

\DeclarePairedDelimiterX\braket[2]{\langle}{\rangle}{#1 \delimsize\vert #2}

\newcommand{\bg}{ \begin{gather} }
\newcommand{\eg}{\end{gather}}
\newcommand{\be}{ \begin{equation} }
\newcommand{\ee}{\end{equation}}
\newcommand{\bea}{ \begin{eqnarray} }
\newcommand{\eea}{\end{eqnarray}}

\renewcommand{\Re}{\mathop{\rm Re}}
\renewcommand{\Im}{\mathop{\rm Im}}

\AtEndEnvironment{thebibliography}{}

\begin{document}

\title{Critical behavior at the localization transition on random regular graphs}

\author{K.\,S.~Tikhonov}
\affiliation{Institut f{\"u}r Nanotechnologie, Karlsruhe Institute of Technology, 76021 Karlsruhe, Germany}
\affiliation{Condensed-Matter Physics Laboratory, National Research University Higher  School of Economics, 101000 Moscow, Russia}
\affiliation{L.\,D.~Landau Institute for Theoretical Physics RAS, 119334 Moscow, Russia}

\author{A.\,D.~Mirlin}
\affiliation{Institut f{\"u}r Nanotechnologie, Karlsruhe Institute of Technology, 76021 Karlsruhe, Germany}
\affiliation{Institut f{\"u}r Theorie der Kondensierten Materie, Karlsruhe Institute of Technology, 76128 Karlsruhe, Germany}
\affiliation{L.\,D.~Landau Institute for Theoretical Physics RAS, 119334 Moscow, Russia}
\affiliation{Petersburg Nuclear Physics Institute,188300 St.\,Petersburg, Russia.}

\begin{abstract}
We study numerically the critical behavior at the localization transition in the Anderson model on infinite Bethe lattice and on random regular graphs. The focus is on the case of coordination number $m+1 = 3$, with a box distribution of disorder and in the middle of the band (energy $E=0$), which is the model most frequently considered in the literature.
As a first step, we carry out an accurate determination of the critical disorder, with the result $W_c =18.17\pm 0.01$. After this, we determine the dependence of the correlation volume $N_\xi = m^\xi$  (where $\xi$ is the associated correlation length) on disorder $W$ on the delocalized side of the transition, $W < W_c$, by means of population dynamics. The asymptotic critical behavior is found to be $\xi \propto (W_c-W)^{-1/2}$, in agreement with analytical prediction. We find very pronounced corrections to scaling, in similarity with models in high spatial dimensionality and with many-body localization transitions.
\end{abstract}
\maketitle

%%%%%%%%%%%%%%%%%%%%%%%%%%%%%%%%%%%%%%%%%%%%%%%%%%%%%%%%%%%%%%%%%%%%%

\section{Introduction}
\label{sec:introduction}

Anderson localization \cite{anderson58} and, in particular, transitions between localized and delocalized phases \cite{evers08} are among central themes of the condensed matter physics. Recently, there was a resurgence of interest to the Anderson models on random regular graphs (RRG) and on related tree-like graphs, largely in view of their relation to the Fock-space representation of interacting problems. Such relations have been pointed out for interacting quantum dot models \cite{altshuler1997quasiparticle, mirlin97,gornyi2016spectral} as well as for many-body localization (MBL) problems with spatially localized single-particle states and short-range \cite{Gornyi2005,Basko2006,gornyi2016spectral} or long-range \cite{gutman2015energy,tikhonov18} interactions. The RRG ensemble is defined as that of random graphs with fixed coordination number  (that is kept constant when one considers the limit of large number of sites $N$). A close relative of the Anderson model on RRG is the sparse random matrix (SRM) ensemble (also known as Erd\"os-R\'enyi graphs in mathematical literature) studied analytically in Refs.~\onlinecite{mirlin1991universality,fyodorov1991localization,fyodorov1992novel}. The central property of the RRG and SRM ensembles is that they represent tree-like models without boundary (and with loops of typical size $\sim \ln N$). The difference between the RRG  and SRM models is that the connectivity is strictly fixed in the first case and is allowed to fluctuate in the second; this difference is inessential for the localization physics that we are interested in.  In view of the connections to many-body problems mentioned above, one can view the Anderson model on RRG as a toy-model of MBL.

It was shown in Refs.~\onlinecite{mirlin1991universality,fyodorov1991localization,fyodorov1992novel} that delocalized states in the SRM model have ergodic properties in the large-$N$ limit. These results were derived by means of a functional-integral representation of the correlation functions of the model in the framework of the supersymmetry formalism. In the large-$N$ limit, the integral can be evaluated by the saddle-point method. The corresponding saddle-point equation has a form analogous to the self-consistency equations obtained for the Anderson model \cite{abou1973selfconsistent,mirlin1991localization}
and the $\sigma$ model \cite{efetov1985anderson,zirnbauer1986localization,zirnbauer1986anderson,efetov1987density,efetov1987anderson,verbaarschot1988graded} on an infinite Bethe lattice.

Numerical verification of the analytic predictions of Refs.~\onlinecite{mirlin1991universality,fyodorov1991localization,fyodorov1992novel} turned out to be not so straightforward. In particular, the works Refs.~\onlinecite{biroli12,deluca14} questioned the ergodicity of the delocalized phase in the RRG model as defined by large-$N$ limit of the energy-level statistics and of scaling of the inverse participation ratio (IPR). Later, the analysis of Ref.~\onlinecite{tikhonov2016anderson} revealed a crossover from relatively small ($N\ll N_\xi$) to large ($N\gg N_\xi$) systems, where  $N_\xi$ is the correlation volume. For $N\ll N_\xi$ the system exhibits  a flow towards the Anderson-transition fixed point which has on RRG properties very similar to the localized phase. When the system volume $N$ exceeds $N_\xi$, the direction of flow is reversed and the system approaches its $N\to\infty$ ergodic behavior. The overall evolution with $N$ is thus non-monotonic. In combination with exponentially large values of the correlation volume $N_\xi$, this makes the finite-size analysis very non-trivial. The main conclusions of Ref.~\onlinecite{tikhonov2016anderson} (in particular, the ergodicity of the delocalized phase)  have been supported by subsequent studies of the IPR scaling in the SRM-like model \cite{garcia-mata17} and of the level number variance in the RRG model \cite{metz17}, as well as in a recent extensive study of the RRG problem  \cite{biroli2018}. Further, we have recently performed \cite{tikhonov19}  a detailed analytical and numerical investigation of the level and eigenfunction statistics on RRG, both in the critical regime ($N\ll N_\xi$) and in the delocalized phase ($N\gg N_\xi$). On the analytical side, we have extended the analysis of Refs.~\onlinecite{mirlin1991universality,fyodorov1991localization,fyodorov1992novel} to the RRG model, in which case the saddle-point equation turns out to be identical to the self-consistency equation \cite{abou1973selfconsistent,mirlin1991localization} for the infinite Bethe lattice, and used it to calculate various observables. We have shown that these predictions, in combination with a numerical solution of the self-consistency equation, are in a perfect agreement with exact-diagonalization results for the eigenfunction and level statistics (in particular, for the IPR in the delocalized phase at $N\gg N_\xi$) on RRG.

As the saddle-point solution determines all physical observables on the RRG, it is important to understand its properties. This solution is intimately related to the distribution function $\mathcal{P}(\Re G, \Im G)$ of the local Green function $G$ on an infinite  Bethe lattice \cite{abou1973selfconsistent,mirlin1991localization}.
The central role for the ergodicity of the extended phase on RRG (disorder $W < W_c$, where $W_c$ is the Anderson-transition point)  is played by the spontaneous symmetry breaking which manifests itself in the emergence of a non-zero typical value of the local density of states (LDOS) $\rho = (1/\pi) \Im G$. Formally, the typical LDOS can be defined as
$\rho_{\rm typ} = \exp \left<\ln\rho\right>$, where $\langle \ldots \rangle$ denotes the disorder averaging. The typical LDOS on the infinite Bethe lattice is directly related to the correlation volume $N_\xi$ that is a key parameter characterizing the self-consistent solution in the delocalized phase, $\rho_{\rm typ} \sim N_\xi^{-1}$ or, equivalently,
\be
\label{ln-nu}
\left<\ln\rho\right> \simeq -\ln N_\xi.
\ee
Moments of $\rho$ on an infinite Bethe lattice are also determined by $N_\xi$; e.g.,
\be
\label{nu2}
\langle \rho^2 \rangle \sim N_\xi.
\ee
As has been emphasized above, the scale $N_\xi$ controls the finite-size scaling of the RRG model: a  not too large system, $N \ll N_\xi$, behaves as critical, while for $N \gg N_\xi$ the ergodicity emerges. As an important example, the average  IPR  $P_2$ of eigenstates is of order  unity for $N \ll N_\xi$ and is equal to
\be
\label{IPR}
P_2 =\frac{3}{N} \frac{\left<\rho^2\right>}{\left<\rho\right>^2}\sim \frac{N_\xi}{N}
\ee
for $N \gg N_\xi$. It is worth emphasizing that the equality in Eq.~(\ref{IPR}) connects the IPR $P_2$ of states in the RRG model with fluctuations $\langle\rho^2\rangle$ of the LDOS in the infinite-Bethe-lattice model \cite{tikhonov19}.

As for any phase transition, one of central questions is that of critical behavior near the transition point $W_c$. On the delocalized side, it is particularly important to know the scaling of the correlation volume $N_\xi$.  As the linear size is proportional to the logarithm of the volume on a Bethe lattice (or RRG), it is natural to expect the scaling
\be
\ln N_\xi \sim (W_c-W)^{-\nu_{\rm del}},
\label{crit-volume-scaling}
\ee
where the subscript ``del'' of the critical index indicates that we deal with the delocalized side of the transition.
Indeed, the analysis of the Anderson model on an infinite Bethe lattice \cite{mirlin1991localization} yielded Eq.~(\ref{crit-volume-scaling}) with
\be
\nu_{\rm del}=1/2,
\label{mu}
\ee
in analogy with earlier results for the $\sigma$ model \cite{zirnbauer1986localization,efetov1987density}.

Numerical data of Refs.~\cite{tikhonov2016anderson,garcia-mata17,biroli2018} for the RRG model were roughly consistent with this prediction, although an accurate determination of the correlation-volume scaling was not in the center of these studies. On the other hand, the value (\ref{mu}) of the critical index $\nu_{\rm del}$ was questioned in Ref.~\cite{kravtsov2018non} where both analytical and numerical arguments in favor of a different value, $\nu_{\rm del}=1$, were put forward.

In the present work, we resolve this question. The main subject of this work is a high-precision numerical analysis of the critical behavior of the scaling of the correlation volume $N_\xi$.
As a first step towards this goal, we accurately evaluate the critical disorder $W_c$ using the approach based on the stability analysis of Ref.~\cite{abou1973selfconsistent}.
To simplify comparison with the previous literature, we choose the model that was studied in most of previous works (coordination number $m+1 = 3$ and box distribution of disorder). By carefully analyzing and eliminating numerical errors, we find  $W_c=18.17$, with an uncertainty less than $0.01$.
A precise knowledge of the critical point is very helpful for an accurate determination of the critical index, since otherwise $W_c$ should be used as an additional fitting parameter. Equipped in addition with large pool sizes (which allows us to approach $W_c$ closely), we are able to firmly establish the value of the critical index, $\nu_{\rm del}=1/2$, from the population-dynamics solution of the self-consistency equation.  We also consider the form of the LDOS distribution $\mathcal{P}(\Im G)$ near criticality and compare it to analytical prediction. In this connection, we also clarify the source of the error in the analytical argumentation of Ref.~\cite{kravtsov2018non} with respect to the value of $\nu_{\rm del}$.

The structure of the paper is as follows. In Sec.~\ref{sec:model} we define the Anderson models on an infinite Bethe lattice and on RRG to which our analysis applies.
The critical disorder $W_c$ is determined numerically in Sec.~\ref{sec:criticalw}. In Sec.~\ref{sec:delocalized} we explore the disorder dependence of correlation volume $N_\xi$
at the delocalized side of the transition and the determine the associated critical index, $\nu_{\rm del}=1/2$. Finally, Sec.~\ref{sec:summary} contains a summary of our results as well as a discussion of their connections to Anderson localization transitions in high spatial dimensionality $d$ and to MBL transitions.

\section{Model}
\label{sec:model}

The analysis of the correlation volume and of the self-consistent solution that we perform in this paper applies to two related models of  non-interacting particles hopping over a graph with fixed connectivity $m+1$  in a potential disorder,
\begin{equation}
\label{H}
\mathcal{H}=\sum_{\left<i, j\right>}\left(c_i^\dagger c_j + c_j^\dagger c_i\right)+\sum_{i=1} \epsilon_i c_i^\dagger c_i\,,
\end{equation}
where the sum is over the nearest-neighbor sites. The energies $\epsilon_i$ are independent random variables sampled from a distribution $\gamma(\epsilon)$ chosen as a box distribution, i.e., a uniform distribution on $[-W/2,W/2]$.
We will study the models at zero energy, $E=0$ (i.e., in the middle of the band).

The first model is that of an infinite Bethe lattice \cite{abou1973selfconsistent,mirlin1991localization}. In this model, the correlation volume $N_\xi$, Eq.~(\ref{crit-volume-scaling}),  characterizes the distribution of LDOS \cite{mirlin1991localization}. To define the corresponding self-consistency equation, one has to introduce a small imaginary part $\eta$ of the energy. The symmetry breaking characterizing the delocalized phase implies that the distribution of LDOS has a non-singular limit at $\eta \to 0$,
which determines the value of the correlation volume under interest, see Eqs.~(\ref{ln-nu}) and (\ref{nu2}). The notion of an infinite Bethe lattice corresponds to the limit $N \to \infty$ taken before the limit $\eta \to 0$.

The second model is the Anderson model on RRG. This model has by definition a finite (although large) number of sites $N$. The observables of interest are statistical properties of eigenfunctions and of energy levels. As has been discussed in Sec.~\ref{sec:introduction}, on the delocalized side of the transition, the critical volume $N_\xi$ marks a crossover from the critical regime, $N \ll N_\xi$, to the ergodic behavior at $N \gg N_\xi$ \cite{fyodorov1991localization,tikhonov2016anderson,garcia-mata17,biroli2018,tikhonov19}.  Furthermore, $N_\xi$ determines various properties of the delocalized phase at $N \gg N_\xi$, including the coefficient in the ergodic ($1/N$) scaling of the IPR, Eq.~(\ref{IPR}), the spatial range of strong correlations in  amplitudes of an eigenfunction,  and the energy-level statistics beyond the random-matrix-theory regime \cite{tikhonov19}.

The self-consistency equation for the infinite Bethe-lattice model (or, equivalently, the saddle-point equation for the RRG model) can be presented in the form
\be
\label{pool_sc}
G^{(m)}\stackrel{d}{=}\frac{1}{E-i\eta-\epsilon-\sum_{i=1}^{m} G_i^{(m)}},
\ee
where the symbol $\stackrel{d}{=}$ denotes the equality in distribution for random variables.
In this equation, $G^{(m)}$ has a meaning of the (advanced) Green function with coinciding spatial arguments, $G^{(m)} = G_{\rm A}(0,0;E) = \langle 0| (E- {\cal H} - i \eta)^{-1}|0\rangle$, defined on a slightly modified lattice, with the site 0 having only $m$ neighbors.  On the right-hand-side of Eq.~(\ref{pool_sc}), $G_i^{(m)}$ are independent, identically distributed copies of the random variable $G^{(m)}$ and $\epsilon$ is a random variable with distribution  $\gamma(\epsilon)$.  Equivalently, Eq.~(\ref{pool_sc}) can be written as a non-linear integral equation for the probability distribution of $G^{(m)}$. We refer the reader to Eq.~(4.6) of Ref.~\cite{abou1973selfconsistent} and to Eqs.~(17), (18) of
Ref.~\cite{mirlin1991localization} for two different representations of this equation; their equivalence is proven in Appendix C of Ref.~\cite{mirlin1991localization}.

Once the self-consistency equation is solved, one can calculate the distribution of the local Green function $G^{(m+1)}$ on an original lattice (with all sites having $m+1$ neighbors) from an auxiliary relation
\be
\label{pool_simple}
G^{(m+1)}\stackrel{d}{=}\frac{1}{E-i\eta-\epsilon-\sum_{i=1}^{m+1} G_i^{(m)}}.
\ee
The distributions of $G^{(m)}$ and $G^{(m+1)}$ have qualitatively very similar properties.
All the information about the position of the transition and the critical volume $N_\xi$ is contained in the self-consistency equation (\ref{pool_sc}).  Below we use short notations $g \equiv G^{(m)},\;G \equiv G^{(m+1)}$. We also focus on the case $m=2$ in the rest of the paper.

%%%%%%%%%%%%%%%%%%%%%
\begin{figure*}[tbp]
\minipage{0.48\textwidth}\includegraphics[width=\textwidth]{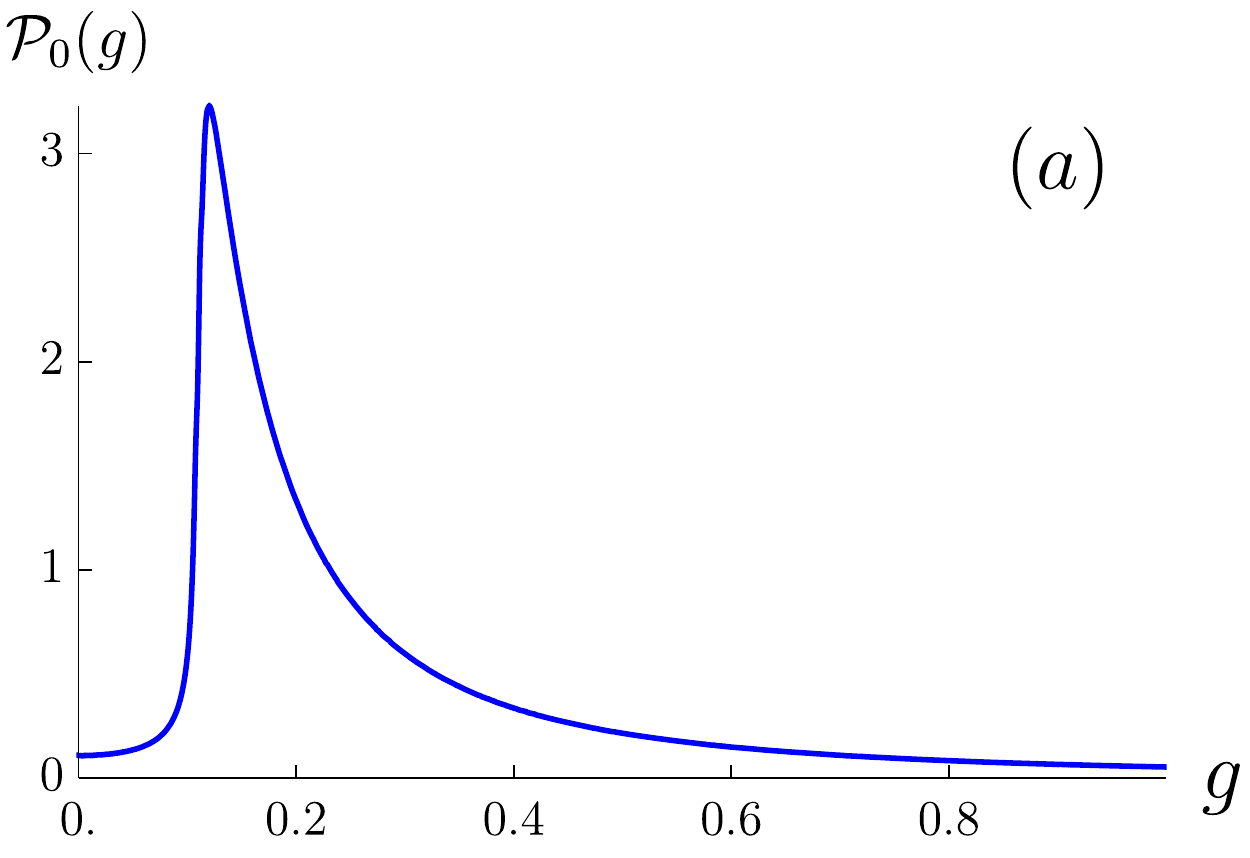}\llap{\makebox[5.5cm][l]{\raisebox{2cm}{\includegraphics[height=3cm]{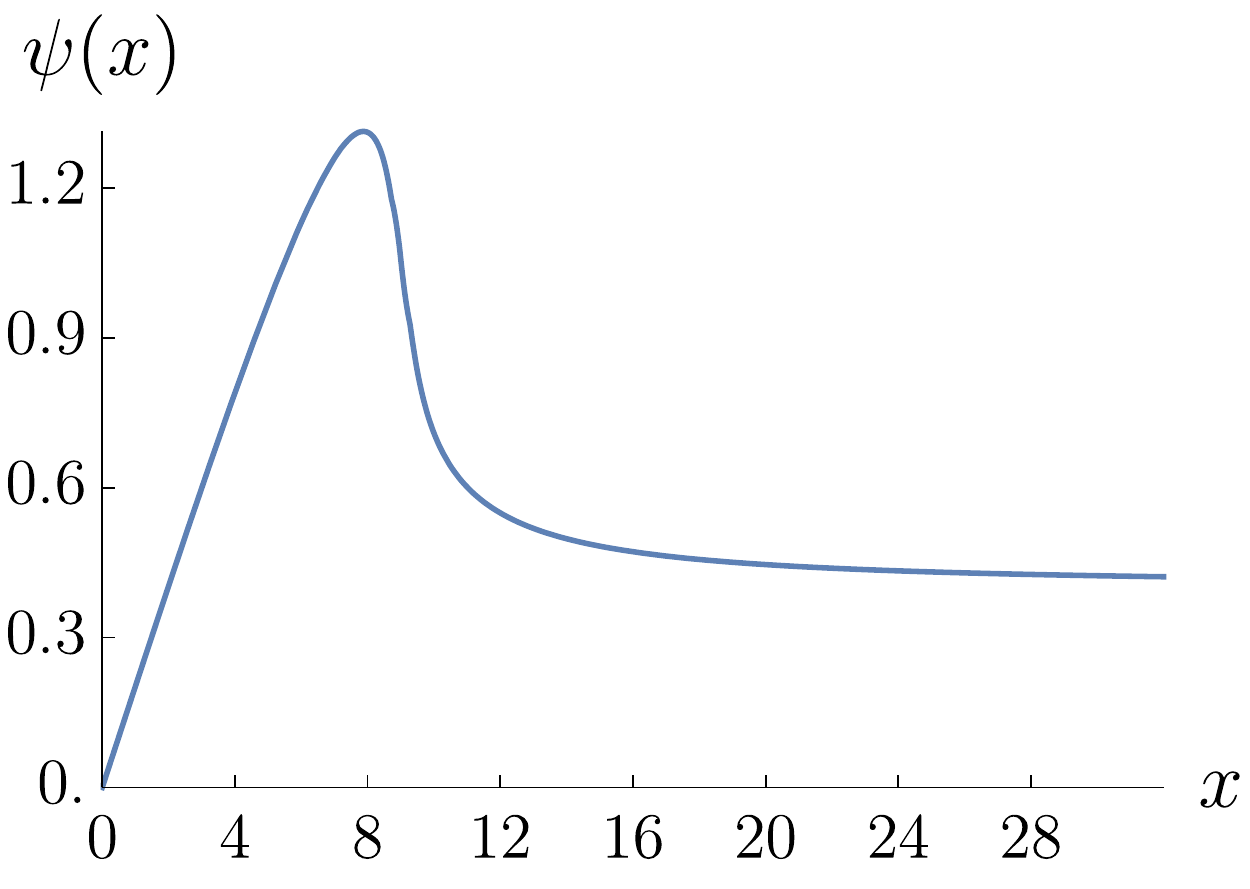}}}}\endminipage
\hspace{0.5cm}
\minipage{0.48\textwidth}\includegraphics[width=\textwidth]{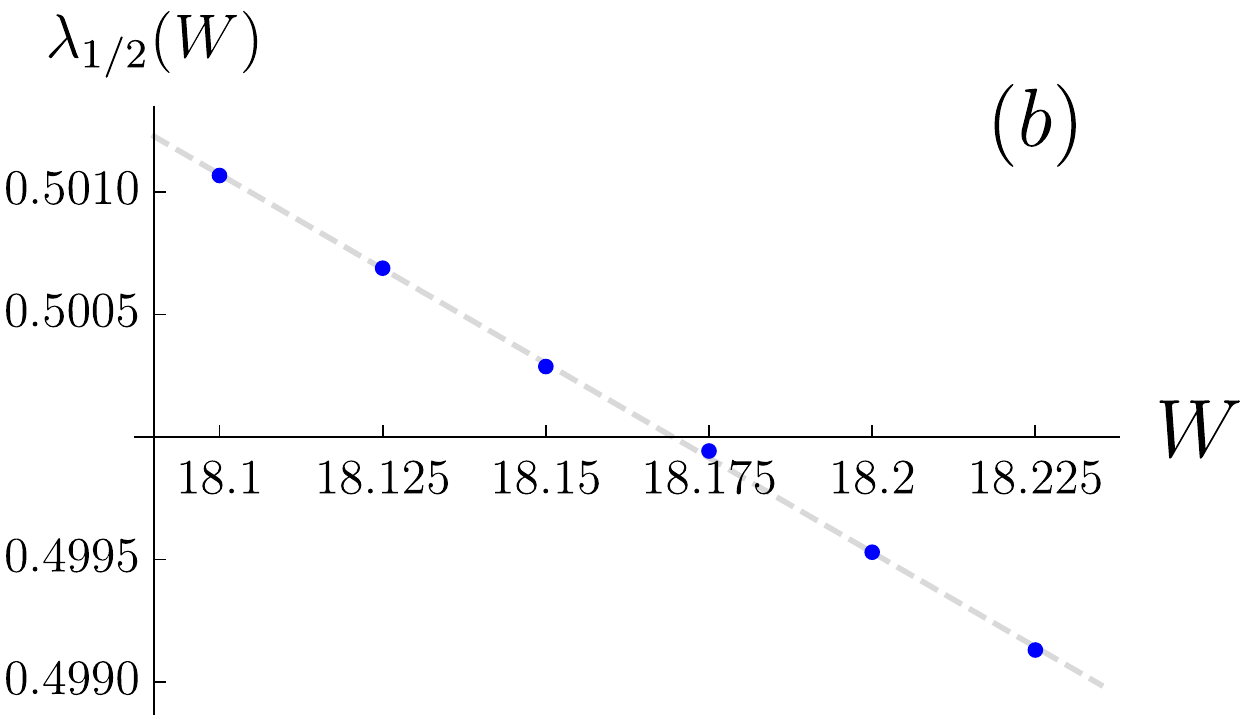}\endminipage
\caption{(a)  Distribution $\mathcal{P}_0(g)$ of the real on-site Green function ($\eta=0$) at $W=18$. Inset: Eigenfunction $\psi(x)$ of the operator in Eq. (\ref{eq:linear2}), corresponding to the largest eigenvalue, at $W=18$. (b) The largest eigenvalue $\lambda_{1/2}(W)$ of Eq. (\ref{eq:linear2}) in the disorder  range $W=18.1-18.225$. The crossing point with the horizontal axis, $\lambda_{1/2} = 1/2$, determines the critical disorder, $W_c = 18.17$.}
\label{fig:pg}
\end{figure*}
%%%%%%%%%%%%%%%%%%%%%%%%

\section{Critical disorder}
\label{sec:criticalw}

In this section, we perform an accurate evaluation of the critical disorder $W_c$ for the model with box distribution and coordination number $m=2$. The strategy of finding the transition point was established in Ref.~\cite{abou1973selfconsistent}; equivalent results were later obtained within the supersymmetry formalism in Ref.~\cite{mirlin1991localization}. Very recently, the same equations were re-derived in Ref.~\cite{parisi2018}. The procedure consists of two steps. First, one discards imaginary parts by setting $\eta =0$ and calculates the resulting distribution $\mathcal{P}_0(g)$ of real on-site Green functions $g$. Then, one investigates the stability of this solution with respect to introducing a small imaginary part. This amounts to evaluating the largest eigenvalue of a certain integral operator, whose kernel involves $\mathcal{P}_0(g)$. The stability (instability)  implies that the system is in the localized (resp., delocalized) phase.

We begin with the first step. The distribution $\mathcal{P}_0(g)$ can be found from Eq.~(\ref{pool_sc}) at $\eta=0$.  The corresponding non-linear integral equation is Eq.~(6.1) of Ref.~\cite{abou1973selfconsistent} and Eq.~(10) [or, equivalently, the unnumbered equation preceding Eq.~(23)] of Ref.~\cite{mirlin1991localization}.

A standard approach for solving equations such as Eq.~(\ref{pool_sc}) is population dynamics, also known as pool method.
With a given pool size $M$, we iterate the self-consistency equation (with $\eta =0$)  until convergence. As a result we get a sample of $M$ variables distributed according to $\mathcal{P}_0(g)$ and recover the distribution function binning the data and interpolating the resulting counts. The result is shown in Fig.~\ref{fig:pg}a for $W=18$; since $\mathcal{P}_0(g)$ is an even  function of $g$, we show it only for $g \ge 0$.  The distribution $\mathcal{P}_0(g)$ has a relatively sharp peak at $g=\pm 2/W$, which is a consequence of the box distribution with the border at the energy $W/2$. Further, very generally, $\mathcal{P}_0(g)$ has a power-law tail, $\mathcal{P}_0(g) \simeq \nu/g^2$ at $g \to \infty$, where $\nu$ is the average DOS.

Having evaluated $\mathcal{P}_0(g)$, we turn to the second step.
Upon introduction of a small level broadening $\eta$ (imaginary part of the energy), the Green function $g$ becomes complex, and
one has to study the joint distribution $\mathcal{P}(\Re g,\Im g)$. The distribution of imaginary part, $\mathcal{P}(\Im g)$, may have either regular or singular limit at $\eta\to 0$. The first case (i.e., spontaneous emergence of a non-trivial distribution of $\Im g$) corresponds to the delocalized phase, the second case to the localized phase.
The transition between these two types of behavior happens at $W_c$, which is a point of the localization transition. In order to find $W_c$, one studies the stability of the
localized phase, i.e., of the distribution $\mathcal{P}_0(g)$ with purely real $g$. The latter is stable if and only if the largest eigenvalue $\lambda_\beta$ of the linear integral operator with the kernel
\be
L_\beta(x,y)=\frac{|x|^{2\beta}}{y^2}\int d\epsilon\, \gamma(\epsilon)\,  \mathcal{P}_0(y^{-1}-x-\epsilon)
\label{eq:linear}
\ee
 is smaller than $1/m$ for $\beta=1/2$. For purpose of brevity, we have written the kernel $L_\beta(x,y)$ in Eq.~(\ref{eq:linear}) for the particular case $m=2$. The general form (valid for arbitrary $m$) of the operator $L_\beta$ can be found in Eq.~(6.5) of Ref.~\cite{abou1973selfconsistent} and in Eq.~(26) of Ref.~\cite{mirlin1991localization}.

One should therefore study the eigenvalue problem
\be
\int L_{1/2}(x,y)\psi(y) dy=\lambda_{1/2}\psi(x).
\label{eq:linear2}
\ee
The spectrum and eigenfunctions of this equation evolve smoothly as functions of disorder. One way to study Eq.~(\ref{eq:linear2}) would be to discretize operator in Eq.~(\ref{eq:linear}) \cite{parisi2018}. However, since the kernel of this equation as well as the solution vary steeply in certain regions of arguments (see Fig.~\ref{fig:pg}), it is very difficult to find the eigenvalue with high accuracy in this manner, unless the mesh is chosen very carefully. We prefer to solve Eq. (\ref{eq:linear}) iteratively,
\be
\psi_{n+1}(x)=\hat L_{1/2} \psi_{n}(x),
\label{eq:iteration}
\ee
choosing the adaptive mesh at each step of the iterative procedure, so that accuracy of the representation is kept above certain predefined threshold. In order to further improve the accuracy, we use the following asymptotic formula:
\be
\psi_{n+1}(x)=\psi_{n}^\prime(0)+ \frac{\rm const}{x},  \qquad x\gg 1,
\label{eq:linear_prop}
\ee
which follows from Eqs.~(\ref{eq:linear}), (\ref{eq:iteration}).
In our iterative scheme, we employ this equation by introducing a parameter $x_{m}$ such that for $x>x_m$ the function $\psi_{n+1}$ is represented by the asymptotic tail
(\ref{eq:linear_prop}) with coefficient chosen to match the known value of $\psi^\prime_n(0)$ to the calculated points at $x<x_m$. The value of $x_m$ is chosen in such a way that the eigenvalue is found with sufficient accuracy (see below).

In the end, we aim to obtain $\lambda_{1/2}$ with 4 significant digits. It turns out that the largest eigenvalue of the operator $\hat L_{1/2}$ strongly dominates all others, so that 6 iterations are sufficient to find this eigenvalue  to the desired accuracy (see Figs. S1a,b in Supplemental Material where we show $\psi_n(x)$ found from repeated application of the recursive equation).  The corresponding eigenfunction for $W=18$ is shown in the inset of Fig. \ref{fig:pg}a.

 In order to find the critical point, we perform the described procedure for several values of $W$ in the range $18.1-18.225$; the results are shown in Fig.~\ref{fig:pg}b. Solving the equation $m\lambda_{1/2}(W)=1$ (with $m=2$), we find
 \be
 \label{Wc}
 W_c=18.17.
 \ee
 To understand the accuracy of the result, it is important to analyze possible sources of errors. Formally, we perform all computations (integrations and interpolations) so that at least 6 digits of the result are expected to be correct. However, the method of the calculation itself has several parameters, finiteness of which is the source of a systematic error: i) the pool size $M$, ii) the histogram bin size $b$, iii) the cutoff $h$ beyond which the function $\mathcal{P}_0(g)$ is replaced by its $\propto 1/g^2$ tail, and iv) the parameter $x_m$ beyond which the asymptotic formula (\ref{eq:linear_prop}) is used. Statistical error of the result is due to fluctuations of the bin counts and is controlled by the number of realizations of the pools (we have from $2^6$ to $2^{13}$ independent samples for $M$ from $2^{30}$ to $2^{20}$). We have evaluated $\lambda_{1/2}(18.175)$ for a set of parameters and found that upon varying them in a wide range \cite{foot-range}, $\lambda_{1/2}$ evolves only in the 5th digit. (In particular, we observe a dependence on the pool size $M$ in the range $10^6-10^9$ which is at least two orders of magnitude weaker than that of Ref. \cite{parisi2018}). The most important source of a systematic error is $x_m$ (see Supplemental Material, Fig. S1c for the corresponding dependence), and in the final calculation we use $x_m=128$. We thus conclude that the value of the $\lambda_{1/2}(W=18.175)$ has an absolute precision of $10^{-4}$. This translates into the uncertainty in  $W_c$ less than $0.007$. Thus, the estimated uncertainty of our result $W_c=18.17$ is within $\pm 0.01$.

%%%%%%%%%%%%%%%%%%%%%
\begin{figure*}[tbp]
\minipage{0.5\textwidth}\includegraphics[width=\textwidth]{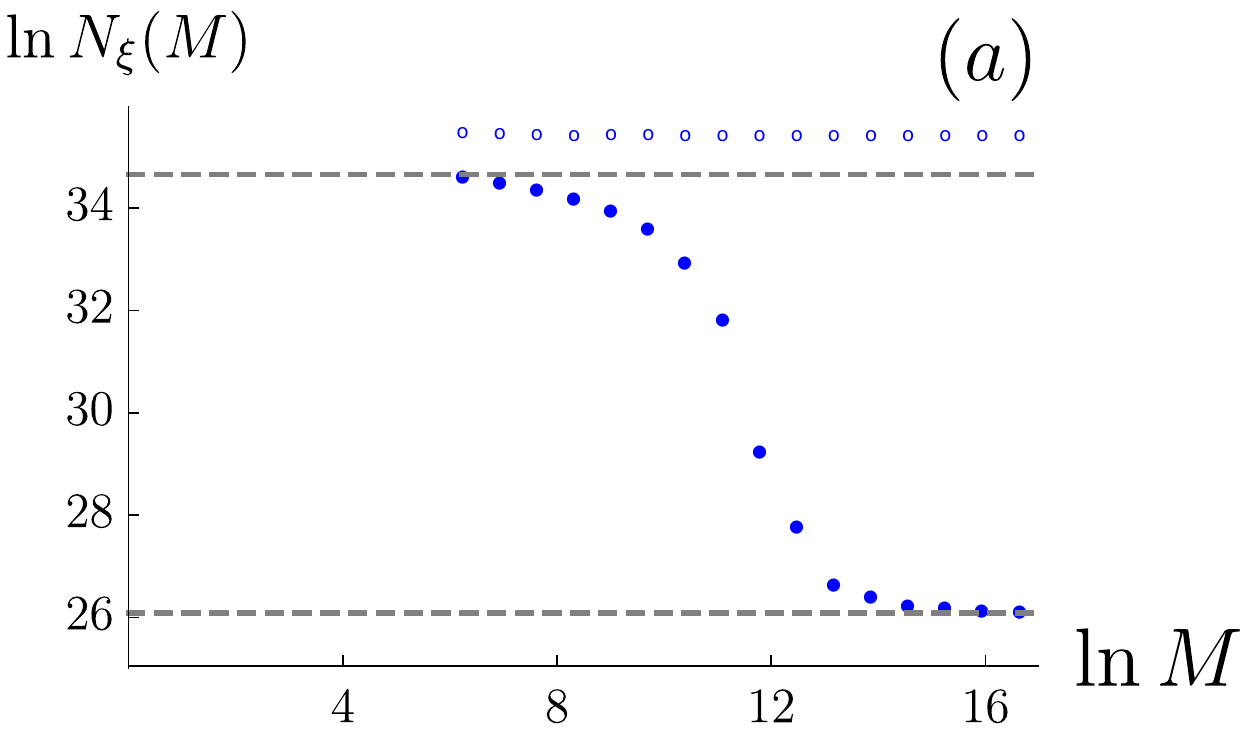}\llap{\makebox[7.8cm][l]{\raisebox{1.2cm}{\includegraphics[height=2.5cm]{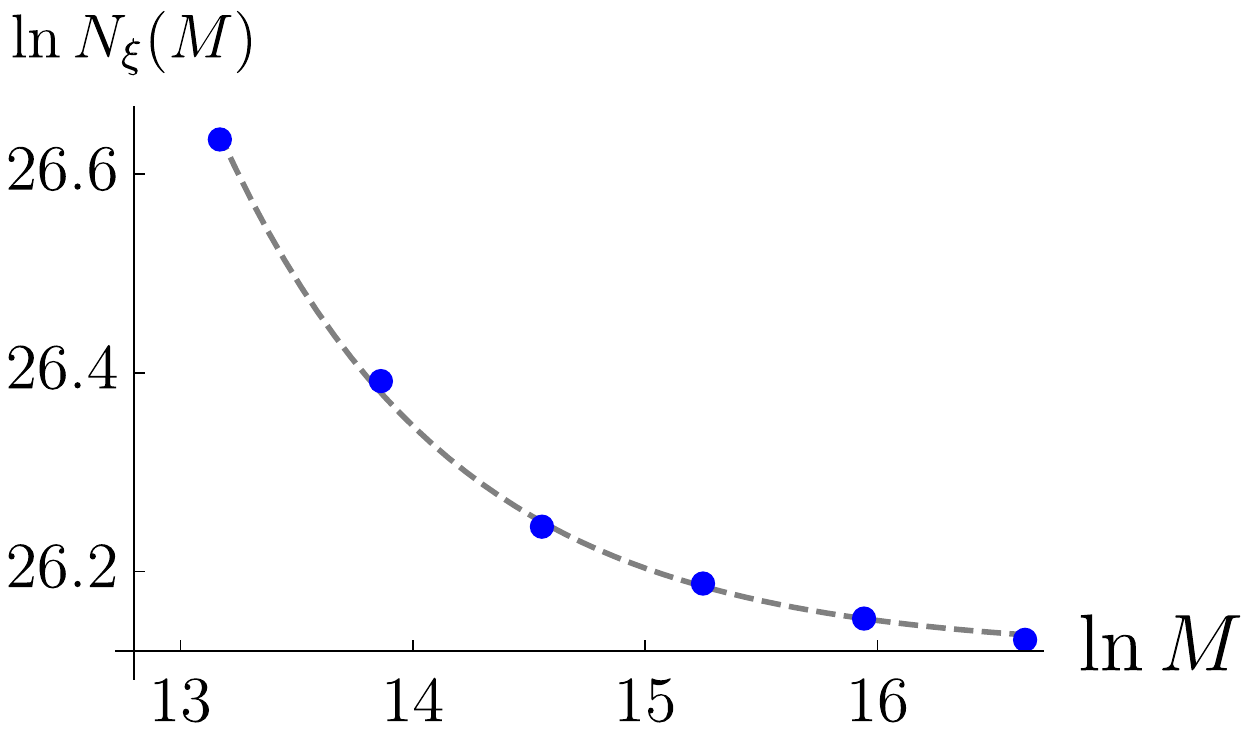}}}}\endminipage
\minipage{0.5\textwidth}\includegraphics[width=\textwidth]{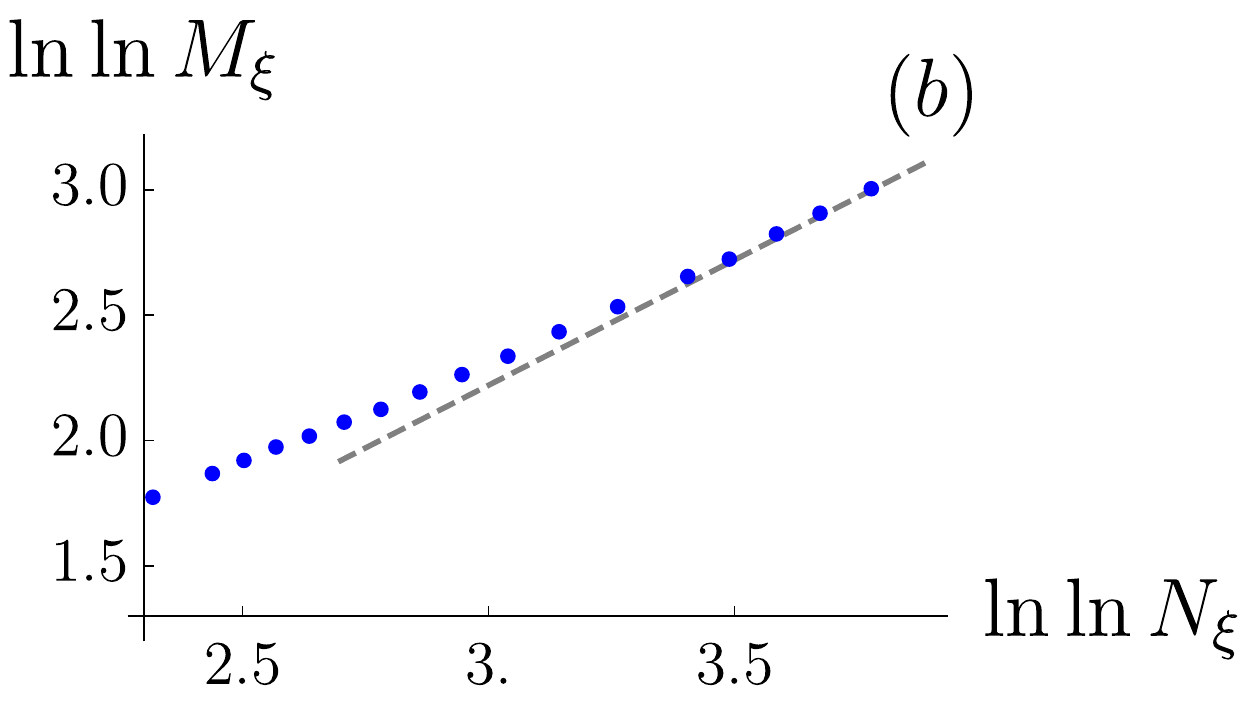}\endminipage
\caption{ (a) Pool size dependence of the correlation volume $N_{\xi}(M)$, defined as population-dynamics result for $\exp\langle - \ln \Im G \rangle$,  at $\ln\eta=-34.6$, $W=17$ (filled symbols) and $W=19$ (empty symbols).  For $W=17$,  $N_{\xi}(M)$ evolves with increasing pool size from $\eta^{-1}$ (upper dashed line) to the true correlation volume $N_\xi \equiv N_{\xi}(M\to\infty)$ (lower dashed line). On the other hand,  for $W=19$ $N_{\xi}(M)$ remains constant of order $\eta^{-1}$. Inset: Zoom of the data for $M\gtrsim M_\xi$, fitted by Eq. (\ref{NxiM}). (b) $M_\xi$, see Eq. (\ref{NxiM}), as function of $N_\xi$. Dashed line: $\ln\ln M_\xi=\ln\ln N_\xi-0.78$.}
\label{fig:mdep}
\end{figure*}
%%%%%%%%%%%%%%%%%%%%%%%%

It is interesting to compare this accurate value of $W_c$ with previous numerical results obtained for the same model. The first such calculation was performed in the pioneering paper \cite{abou1973selfconsistent} and yielded an estimate $W_c \simeq 16$. Subsequent works yielded improved estimates  $W_c \simeq 17.32$ \cite{monthus2008anderson}, and
$W_c \simeq 17.4$ \cite{biroli2010anderson} (improvements were mostly related to increasing the pool size). Recently, three papers aimed for a more accurate determination of $W_c$ and found $W_c=18.2$ \cite{biroli2018}, $W_c=18.8$ \cite{kravtsov2018non}, and $W_c=18.45$ \cite{parisi2018}. As we will demonstrate in the next section, such deviations are actually very strong in the context of determination of the critical behavior, so it is important to understand the origin of the controversies. The value found in Ref.~\cite{biroli2018} has larger uncertainty than our result, and within this uncertainty it is consistent with our Eq.~(\ref{Wc}). The values found in Ref. \cite{parisi2018} and especially in Ref. \cite{kravtsov2018non} deviate more sizeably from our result.  The analysis of Ref. \cite{parisi2018} that yielded $W_c=18.45$ has probably suffered from a not sufficiently accurate solution of Eq. (\ref{eq:linear2}) caused by fixed mesh discretization.
As to Ref. \cite{kravtsov2018non}, we believe that errors in numerical determination of $W_c = 18.8$ resulted from a combination of several sources. First, contrary to our approach, in which we  first determine $W_c$ with high accuracy and then study the critical behavior, the authors of Ref.  \cite{kravtsov2018non} attempted to do this simultaneously, which is less favorable from the point of view of the accuracy in the presence of large corrections to scaling [see discussion after Eq. (\ref{corr-vol-large-m})]. Second, the corresponding analysis was not carried out in Ref. \cite{kravtsov2018non} in an optimal way (see Fig. S4 and the associated comments in Supplemental Material). Finally,
determination of $N_{\xi}$ in this work appears to be plagued by sizeable numerical errors. Apparently, the authors relied on points derived for insufficiently small $\eta$ too close to criticality: the smallest value of $\eta$ that is quoted in the caption to Fig. 8 of Ref. \cite{kravtsov2018non}, $\eta = 10^{-10}$, is way too large, as the studied correlation volumes exceed $10^{10}$ at $W>16.75$.

\section{Delocalized phase}
\label{sec:delocalized}

%%%%%%%%%%%%%%%%%%%%%
\begin{figure*}[tbp]
\minipage{0.5\textwidth}\includegraphics[width=\textwidth]{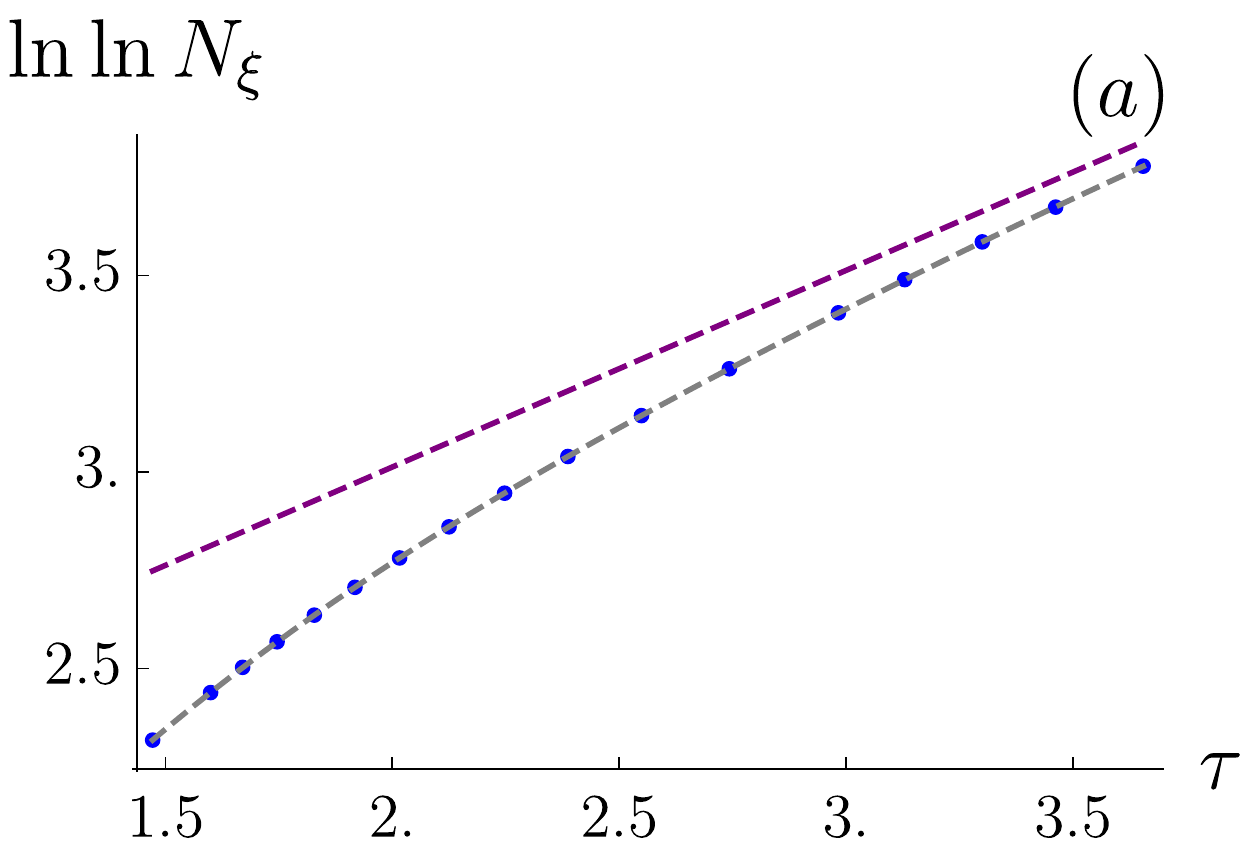}\llap{\makebox[4.5cm][l]{\raisebox{1.2cm}{\includegraphics[height=3cm]{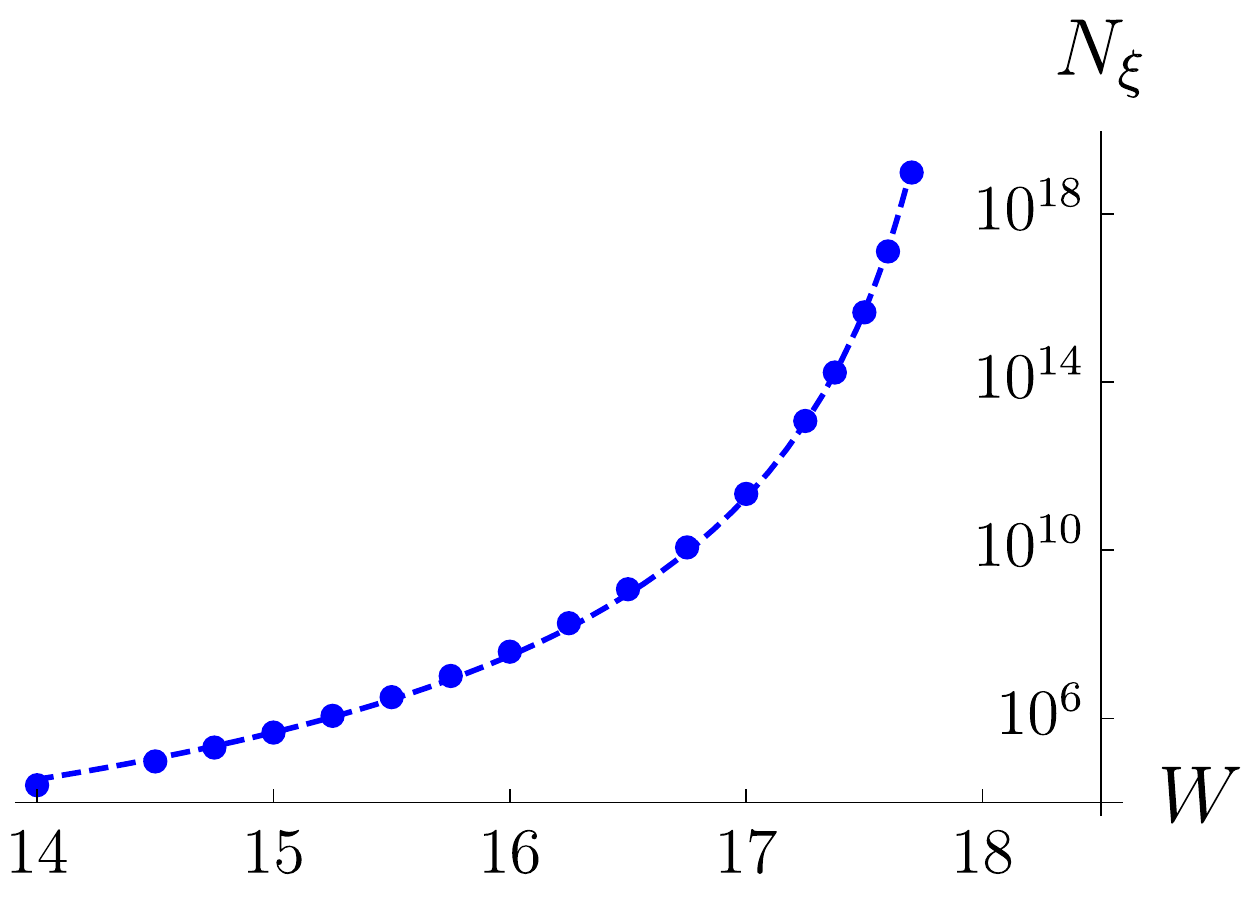}}}}\endminipage
\minipage{0.5\textwidth}\includegraphics[width=\textwidth]{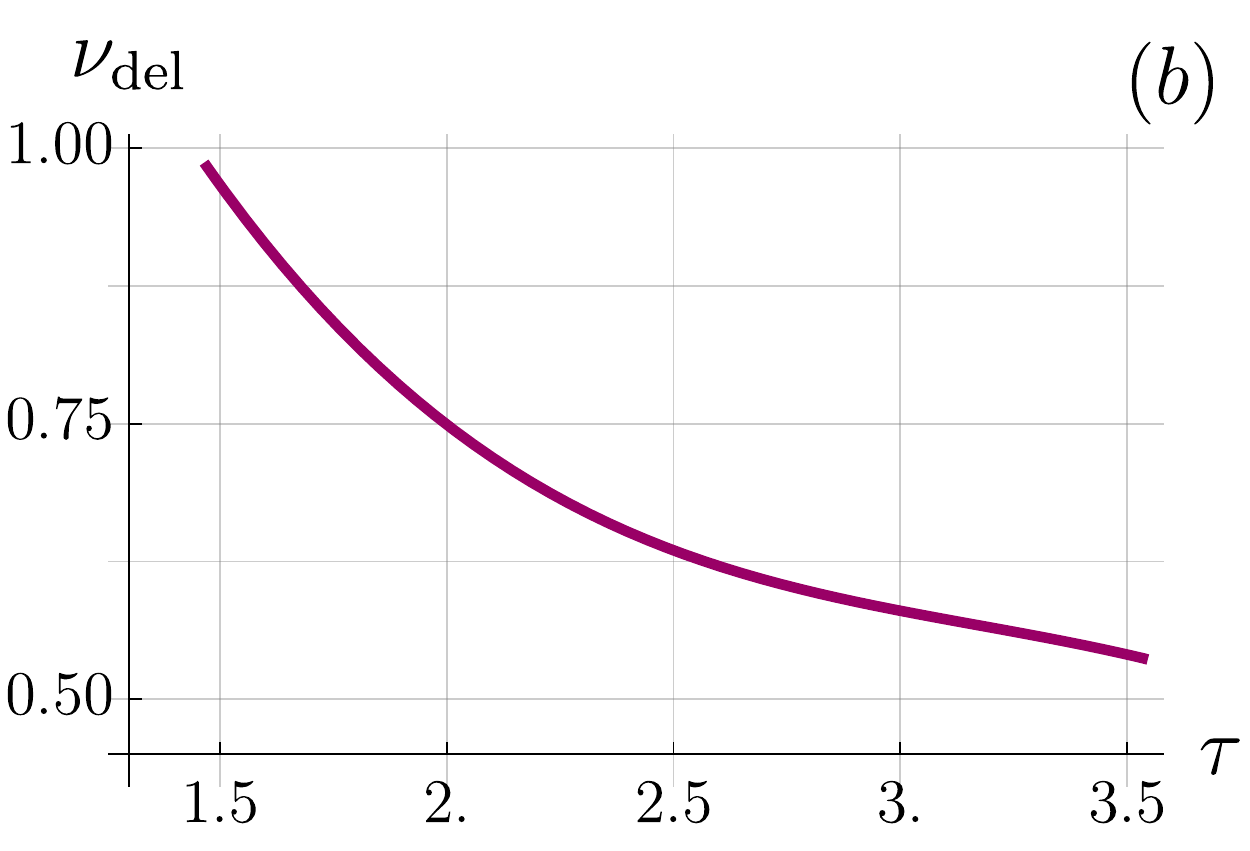}\endminipage
\caption{(a) Double logarithm of the correlation volume $N_\xi$ as a function of $\tau=-\ln(1-W/W_c)$, where $W$ is the disorder strength, and $W_c=18.17$ is the critical point; magenta dashed line: asymptotic behavior, Eq.~(\ref{cvolth}). Inset: the same data in $(W,\,\ln N_\xi)$ coordinates; blue dashed line: Eq. (1) of Supplemental Material, representing the asymptotic behavior (\ref{cvolth}) with included subleading correction.  (b) Flowing correlation-length exponent $\nu_{\rm del}(\tau)=\partial \ln \ln N_\xi/\partial\tau$ as a function of $\tau=-\ln(1-W/W_c)$ with the critical disorder $W_c=18.17$.
The saturation value, $\nu_{\rm del}(\tau \to \infty)$, yields the critical index $\nu_{\rm del} = 1/2$ of the correlation length, Eq.~(\ref{xi}).}
\label{fig:cvol}
\end{figure*}
%%%%%%%%%%%%%%%%%%%%%%%%

With an accurate value of $W_c$ at hand, see Eq. (\ref{Wc}), we turn to the analysis of the critical behavior in the delocalized phase.
As has been already discussed above, the $\eta=0$ solution is unstable at $W<W_c$ to introduction of finite $\eta$.
In the field-theoretical description of the Anderson localization, this phenomenon can be cast in the framework of  spontaneous symmetry breaking, with an order-parameter function intimately related to the distribution of LDOS. As a result, a non-trivial distribution $\mathcal{P}(\Re G, \Im G)$ emerges, which can be found from a complex-valued pool method applied to Eqs.~(\ref{pool_sc}),~(\ref{pool_simple}).

 The population dynamics calculation is conducted at a finite (although large) pool size $M$ and a finite (although small) imaginary part of the energy, $\eta$. The resulting distribution can be characterized by ``effective correlation volume'' $N_\xi(\eta,M)$ defined as the population-dynamics result for $\exp \langle - \ln \Im G \rangle$, where $\langle .. \rangle$ includes averaging over many iterations after the approximate convergence is reached. It is obvious that at finite $M$ the distribution of $G$ sampled by iterative procedure is not exactly that of $M\to\infty$ limit. One can expect that for smaller pool sizes the sample-to-sample fluctuations of the typical value of $\Im G$ are large and decrease with growing $M$. This is indeed the case, see Fig. S2 of Supplemental Material for illustration of the evolution of the iterative procedure with increasing $M$. As soon as $M$ is large enough, these fluctuations mostly average out and the result is weakly dependent on $M$. In the end, the true correlation volume $N_\xi$ is given by the double limit
\be
\label{N-xi-true}
N_\xi=\lim_{\eta\to0}\lim_{M\to\infty}N_\xi(\eta,M).
\ee

To illustrate the role of the pool size, we show in Fig. \ref{fig:mdep}a the dependence of $N_\xi(\eta,M)$ on $M$ for $W=17$ and fixed $\eta = \exp(-34.6)$.
It is seen that for relatively small pool sizes $N_\xi(\eta,M)$ is essentially equal to $\eta^{-1}$, which is a characteristic behavior for the localized phase and the critical point. With increasing $M$, the system ``recognizes'' that it is actually on the delocalized side of the transition, and $N_\xi(\eta,M)$ crosses over to the value $N_\xi(\eta,\infty)$. If $\eta$  has been chosen to be sufficiently small, $\eta \ll N_\xi^{-1}$, the behavior of $N_\xi(\eta,M)$ at large $M$ is essentially independent on  $\eta$, i.e., $N_\xi(\eta,M) = N_\xi (M)$ (see Fig. S3b in Supplemental Material for more detail).  The limiting value $N_\xi (M \to \infty)$ yields the sought physical correlation volume $N_\xi$. This behavior in the delocalized phase should be contrasted to the data for the localized side of the transition, $W=19$, which illustrate the stability of the localized phase with respect to $\eta$.

We find that the large-$M$ approach of $N_\xi(M)$ to its asymptotic value is rather fast and is very well fitted by the $1/M$ dependence:
\be
\label{NxiM}
\ln N_\xi(M) = \ln N_\xi(\infty) + \frac{M_\xi}{M},
\ee
see the inset of Fig.~\ref{fig:mdep}a as well as Fig. S3a of Supplemental Material. We obtain this asymptotic behavior for all studied values of $W$. The parameter $M_\xi$ in Eq.~(\ref{NxiM}) defines the pool size that is needed to obtain $N_\xi$ up to a factor of order unity. In Fig.~\ref{fig:mdep}b we have shown $M_\xi$ as a function of $N_\xi$, with a double-logarithmic scale on each axis. A natural expectation would be that $M_\xi$ is of the order $N_\xi$. Indeed, we observe a unit asymptotic slope of the dependence in Fig.~\ref{fig:mdep}b, which implies that $M_\xi$ has the same asymptotic behavior (\ref{crit-volume-scaling}) as $N_\xi$, with the same exponent $\nu_{\rm del}=1/2$.  On the other hand, there is also a non-trivial intercept,
\be
\label{MN}
\ln \ln M_\xi \simeq \ln \ln N_\xi - c, \qquad c \simeq 0.78,
\ee
which means that
\be
M_\xi \sim N_\xi^\kappa
\ee
with $\kappa \simeq 0.46$. In other words, the pool size needed to find accurately $N_\xi$ is in fact much smaller than $N_\xi$, although it diverges at $W\to W_c$ in the same exponential fashion.

The disorder dependence of the correlation volume $N_\xi$ obtained according to Eq.~(\ref{N-xi-true}) is shown in Fig. \ref{fig:cvol}a.  More specifically, we plot $\ln\ln N_\xi$ as a function of $\tau\equiv-\ln(1-W/W_c)$.  In this representation, the asymptotic slope $\nu_{\rm del}(\infty)$,
where
\be
\nu_{\rm del} (\tau) = \frac{\partial \ln \ln N_\xi}{\partial \tau},
\ee
yields the sought index $\nu_{\rm del}$ in the critical behavior (\ref{crit-volume-scaling}) of the correlation volume. In Fig.~\ref{fig:cvol}b, we show the $\tau$ dependence of the slope $\nu_{\rm del} (\tau)$, which can be viewed as a ``flowing critical index''. It is seen that $\nu_{\rm del}(\tau)$ decreases substantially with increasing $\tau$ and saturates at $\tau \to \infty$. The found saturation value is in a perfect agreement with the analytical  prediction $\nu_{\rm del}=1/2$, Eq.~(\ref{mu}).
We thus have numerically determined the critical behavior of the correlation volume $N_\xi$, or, equivalently, of the corresponding correlation length $\xi$,
\be
\label{xi}
\xi = \frac{\ln N_\xi}{\ln m} \propto \left(W_c-W\right)^{-1/2}.
\ee

In fact, the analytical theory predicts not only the critical index $\nu_{\rm del} = 1/2$ of the scaling (\ref{xi}) of $\ln N_\xi$ but also the corresponding numerical prefactor.
The analysis of the symmetry-broken solution near $W_c$ in the Anderson model \cite{mirlin1991localization} model bears 
(in analogy with its $\sigma$ model counterpart \cite{zirnbauer1986localization,efetov1987density})
a close connection to the analysis of stability of the localized phase.  It leads to the equation 
\be
\label{stability-deloc}
m\lambda_\beta = 1,
\ee 
where $\lambda_\beta$ is the largest eigenvalue of the operator (\ref{eq:linear}).   We recall that  $m=2$ in our case. 
 Expanding the eigenvalue $\lambda_\beta$ around $W = W_c$ and $\beta = 1/2$ up to the leading non-vanishing terms, we get
\be
\lambda_{\beta}(W) \simeq \frac12 - c_1\left(W-W_c\right)+c_2\left(\beta-\frac12\right)^2.
\label{eigval}
\ee
For $W$ below $W_c$,  the solution of Eqs.~(\ref{stability-deloc}), (\ref{eigval}) is
\be
\beta = \frac{1}{2} \pm i\sigma\,, \qquad \sigma \simeq \sqrt\frac{c_1}{c_2}(W_c-W)^{1/2}.
\label{sigma}
\ee 
 The imaginary part $\sigma$ determines, via the equation \cite{mirlin1991localization}
\be
\ln N_\xi \simeq \frac{\pi}{\sigma} \,, 
\label{N-xi-sigma}
\ee
 the correlation volume $N_\xi$. Evaluating numerically the largest eigenvalue $\lambda_\beta$ of operator in Eq. (\ref{eq:linear}) for $\beta$ close to 1/2 and $W$ close to $W_c$, we find 
 \be
 \label{cs}
 c_1\simeq 1.59\textrm{ and }c_2\simeq 0.0154.
 \ee  Substituting these values in Eqs.~(\ref{sigma}), (\ref{N-xi-sigma}), we finally get the numerical value of the prefactor in the asymptotic scaling law for $\ln N_\xi$,
\be
\label{cvolth}
\ln N_\xi \simeq \frac{31.9}{\sqrt{W_c-W}}.
\ee
This asymptotic expression is shown in Fig. \ref{fig:cvol}a by magenta dashed line. The agreement between the population-dynamics results and this analytical prediction for the asymptotic behavior is very impressive. 

It is instructive to compare these exact results with the large-$W_c$ approximation. This approximation is formally valid for $m\gg 1$ (since $W_c$ is proportional to $m\ln m$ for large $m$); we will see, however, that it works very well already for $m=2$ (which is related to the numerically quite large value $W_c=18.17$).  For large $W$, the eigenvalue 
$\lambda_{\beta}$ is given by the asymptotic formula \cite{tikhonov2016fractality}
 \be
\label{abou}
\lambda_{\beta}\simeq \frac{1}{\beta-1/2}\frac{1}{W-4/W}\left[\left(\frac{W}{2}\right)^{2\beta-1}-\left(\frac{W}{2}\right)^{-2\beta+1}\right].
\ee
Substituting this formula in Eq.~(\ref{stability-deloc}) with $\beta=1/2$, one finds the corresponding approximation for the critical disorder $W_c$.  For $m=2$, this gives $W_c\approx 17.65$, which differs only by $3\%$ from the exact result $W_c=18.17$.  Further, expanding Eq.~(\ref{abou}) with respect to $\beta-1/2$ and $W_c-W$, see Eq.~(\ref{eigval}),  and substituting the corresponding coefficients $c_1$ and $c_2$ in Eqs.~(\ref{sigma}), (\ref{N-xi-sigma}), we get\cite{mf}
\be
 \ln N_\xi \simeq \pi\sqrt\frac23 \frac{\ln^{3/2}\left(W_c/2\right)}{\ln^{1/2}\left(W_c/2e\right)}\sqrt{\frac{W_c}{W_c-W}}.
 \label{corr-vol-large-m}
 \ee
For $m=2$, this equation yields $\ln N_\xi \simeq 31.9/\sqrt{W_c-W}$, thus reproducing the correct numerical coefficient in Eq.~(\ref{cvolth}) with amazing accuracy of $0.3\%$.

Let us emphasize the following important point. The fact that $\nu_{\rm del}(\tau)$ saturates at a non-trivial value (actual critical index $\nu_{\rm del}$) at $\tau \to \infty$ is a consequence of the correct choice of the critical point, $W_c=18.17$, in the definition of $\tau$.  If we would use a lower value of $W_c$, the resulting dependence $\nu_{\rm del}(\tau)$ would tend to zero at $\tau \to \infty$, since we would reach infinite $\tau$ while being still in the delocalized phase.  By similar token, taking $W_c$ larger than the actual value, we would get increasing $\nu_{\rm del}(\tau)$ tending to diverge at a finite $\tau$.  This is illustrated in Fig. S4 of Supplemental Material, where we compare the correct $\nu_{\rm del}(\tau)$  ($W_c= 18.17$) with dependencies obtained by using smaller ($W_c = 17.9$, $W_c=18.0$) and larger ($W_c = 18.35$, $18.45$ and $18.8$) values of tentative $W_c$ in the definition of $\tau$. The expected behavior is clearly seen, so that, solely on the basis of the data for $N_\xi$,  we could conclude that $W_c$ is in the range $18.0$ -- $18.35$. (Note that the value $18.8$ proposed in Ref. \cite{kravtsov2018non} is well outside this range; see Supplemental Material for more comments on a deficiency of the numerical procedure in that work.)
 As has been already emphasized, the accuracy of this method of determination of $W_c$ is significantly lower than of that based on investigation of stability of the localized phase, Sec.~\ref{sec:criticalw}. At the same time, the agreement between the two approaches is encouraging.

{The population-dynamics analysis provides not only $N_\xi$ but also the full distribution of the Green function $G$.
In Fig.~\ref{fig:im} we illustrate the distribution of the imaginary part $\mathcal{P}(\Im G)$; this is essentially the distribution of LDOS $\rho$. As expected, we observe the power-law distribution
\be
\label{Pnu}
\mathcal{P}(\rho) \sim N_\xi^{-1/2} \rho^{-3/2}, \qquad N_\xi^{-1}  < \rho < N_\xi;
\ee
outside of this range the probability is strongly suppressed.  This behavior of $\mathcal{P}(\rho)$ is essentially the same as the one found in the $\sigma$ model on the Bethe lattice \cite{mirlin94a,mirlin94b}.  
The LDOS distribution  (\ref{Pnu}) is intimately related to the eigenvalue $\beta$ in Eq.~ (\ref{sigma}). 
Specifically,  the real part 1/2 of $\beta$ translates into the exponent 3/2 in the LDOS distribution (\ref{Pnu}), while the imaginary part determines, via the equation (\ref{N-xi-sigma}),  the correlation volume $N_\xi$ which controls the range of validity of the power-law distribution (\ref{Pnu}).

At this point, it is appropriate to comment on the error in the analytical argument of Ref.~\cite{kravtsov2018non} that suggested an incorrect value of the exponent, $\nu_{\rm del}=1$.
The authors of this paper did not appreciate a difference in the LDOS statistics in the models of infinite and finite Bethe lattices. (In the first the case the limit $N\to \infty$ is taken before the limit $\eta\to 0$; in the second case the order of limits is opposite.)  This difference has been demonstrated in great detail in
Ref.~\cite{tikhonov2016fractality} where the statistics of eigenfunctions and LDOS at a root of the finite Bethe lattice was studied. In particular, the parameter $m_0(W)$ of Ref.~\cite{kravtsov2018non} is the parameter $\beta_*$ of Ref.~\cite{tikhonov2016fractality} that characterizes the fractal statistics on a finite Bethe lattice.
This parameter changes linearly as a function of disorder near $W_c$, which apparently leads to the value $\nu_{\rm del} = 1$ deduced by Ref.~\cite{kravtsov2018non}, see Eqs.~(56)-(60) of that work. The argument is incorrect since the parameter $\beta_*$ of Ref.~\cite{tikhonov2016fractality} (i.e., $m_0(W)$ of Ref.~\cite{kravtsov2018non}) applies  to a finite Bethe lattice but not to the infinite one. This error is closely related to a more general deficiency of Ref.~\cite{kravtsov2018non} that fails to properly discriminate between the fractal properties on a finite Bethe lattice and the ergodicity of the delocalized phase on RRG.

%%%%%%%%%%%%%%%%%%%%%%%%%%%%
\begin{figure}[tbp]
\includegraphics[width=0.5\textwidth]{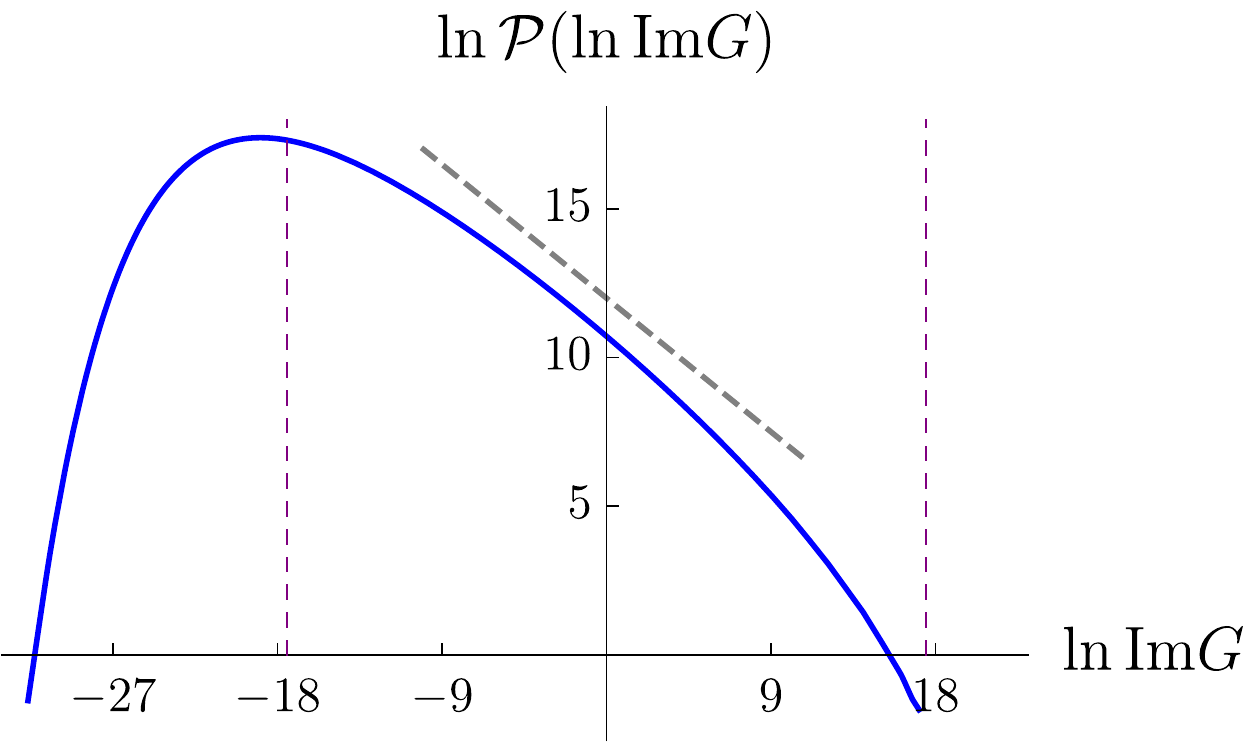}
\caption{Solid: distribution function of the LDOS at $W=16$ as found from the self-consistency equation. Gray dashed line corresponds to $\mathcal P(\Im G)\propto \left(\Im G\right)^{-3/2}$. Vertical dashed lines denote $\pm\ln N_\xi$.}
\label{fig:im}
\end{figure}
%%%%%%%%%%%%%%%%%%%%%%%%%%%%

It is worth emphasizing that the critical index $\nu_{\rm del} = 1/2$ that we have studied in this paper characterizes the correlation length $\xi$ on the delocalized side of the transition, $W < W_c$.  The counterpart of $\xi$ on the localized side, $W > W_c$,  is the localization length $\xi_{\rm loc}$ that is a length controlling the exponential decay of the density-density correlation function (multiplied by the factor $m^r$) with the distance $r$.  This length is given by \cite{mirlin1991localization}
\be
\xi_{\rm loc} = \ln \frac{1}{m\lambda_{1/2}},
\ee
and scales near the critical point as
\be
\xi_{\rm loc} \propto (W-W_c)^{-\nu_{\rm loc}}, \qquad \nu_{\rm loc} = 1,
\ee
in analogy with the corresponding results for the $\sigma$ model \cite{zirnbauer1986localization,efetov1987density}.
The different scaling of the characteristic lengths on both sides of the transition is a special feature of the Bethe-lattice and RRG models, distinguishing them from the conventional $d < \infty$ dimensional models.
We also mention that, on the delocalized side of the transition, further two lengths (much larger than $\xi$) were identified that control the asymptotic decay of the connected part of a LDOS-LDOS correlation function and scale with indices 1 and 3/2 \cite{zirnbauer1986localization,efetov1992scaling}. These lengths appear to be, however, of minor physical importance for the physics of the RRG model. Indeed, it is the length $\xi$ (or, equivalently, the correlation volume $N_\xi$) that determines the finite-size crossover from the critical regime to the ergodic behavior on RRG, thus controlling the associated scaling properties of wave-function and energy level statistics and of further related observables \cite{tikhonov19}.

\section{Summary}
\label{sec:summary}

To summarize, we have studied numerically the critical behavior at the localization transition in the Anderson model on infinite Bethe lattice and on RRG.  We have focused on the case of coordination number $m+1 = 3$, with a box distribution of disorder and in the middle of the band, $E=0$, which is the model most frequently considered in the literature. As a first step, we have carried out an accurate determination of the critical disorder $W_c$, carefully analyzing all essential sources of numerical errors. The resulting value is $18.17\pm 0.01$. After this, we have determined the dependence of the correlation volume $N_\xi$  on disorder $W$ on the delocalized side of the transition, $W < W_c$.  This analysis was done by means of population dynamics, with pool sizes $M$ up to $10^{10}$, and with $N_\xi$ obtained as $\exp \langle - \ln \Im G \rangle$, where $G$ is the one-site Green function and $\Im G$ is the LDOS  (times $\pi$).
Also in this part of the study, we have carefully analyzed convergence with respect to all relevant parameters, including the pool size, the number of iterations, and the imaginary part $\eta$ of the energy.  The resulting dependence $N_{\xi}(W)$ is shown in Fig.~\ref{fig:cvol}a. In view of the relation between the RRG model and that on the infinite Bethe lattice, which was established analytically \cite{mirlin1991universality,fyodorov1991localization,tikhonov19} and confirmed numerically \cite{tikhonov19}, the obtained correlation volume $N_\xi$ also characterizes the RRG problem in which it controls a crossover from criticality to ergodicity.

It is worth emphasizing that we were able to reach controllably values of the correlation volume $N_\xi$ as large as $10^{19}$, which is many orders of magnitudes larger than the volume of a  system that one could study directly (via exact diagonalization). This is because the problem allows to use the population-dynamics approach. First, the maximal pool size $M$ that we employ is $10^{10}$, which is already much larger than the size of a system that can be diagonalized. In addition to this, and quite remarkably, the correlation volume that can be reached turns out to be much larger than the pool size $M$, see Fig.~\ref{fig:mdep}b and Eq.~(\ref{MN}).

With the accurate value of $W_c$ and the dependence of $N_\xi(W)$ at hand, we have determined numerically the critical index $\nu_{\rm del}$ of the correlation length $\xi = \log_m N_\xi$.  For this purpose, we have plotted the flowing exponent $\nu_{\rm del}(\tau)$, see Fig.~\ref{fig:cvol}b. The true exponent $\nu_{\rm del}$ is given by the limit $\nu_{\rm del}(\tau \to \infty)$. The numerically established $\nu_{\rm del}(\tau)$ decreases monotonically and saturates at large $\tau$. The saturation value is in a good agreement with the analytical prediction $\nu_{\rm del} = 1/2$,  Eq.~(\ref{xi}). A substantial (factor-of-two) variation of the ``running exponent'' $\nu_{\rm del}(\tau)$ in Fig.~\ref{fig:mdep} serves as an indication of rather appreciable corrections to scaling, which underlines the importance of proceeding within our analysis up to such large $N_\xi$.

Numerical evaluation of critical indices for non-interacting Anderson transitions was a subject of very intense activity over a few decades. As an example, let us focus on the ``standard model'' of the Anderson transition---that for a 3D system in the orthogonal universality class. Development of the finite-size-scaling approach has allowed one to determine the exponent $\nu$ of the localization (correlation) length as $\nu \simeq 1.5$ \cite{mackinnon1983mackinnon}. Subsequent works have increased the accuracy  and rederived the exponent $\nu$ by several complementary approaches \cite{varga1995varga,zharekeshev1997ik,slevin1999k,milde2000f,rodriguez2010critical,slevin2014critical,slevin2018critical}. In particular, a recent result\cite{slevin2018critical} is $\nu = 1.572 \pm 0.003$, which means an amazingly high precision of $0.2\%$ in determination of the critical index.  One thus might be surprised that a comparable numerical analysis of scaling has not been done long ago also for Anderson models on the Bethe lattice and on RRG.
This is related to a much higher computational complexity of the RRG model in comparison to its 3D counterpart. The reasons for this are twofold. First, the volume $N$ on RRG (or on the Bethe lattice) is an exponential function of length $L$, $N = m^L$.
As an illustration, the huge correlation volume $10^{19}$ (our largest value) on RRG corresponds to the correlation length $\xi =63$ that does not look so large from the point of view of a 3D system.  Second, corrections to a ``simple'' one-parameter scaling are much more pronounced for RRG than for a 3D Anderson transition. As a manifestation of this, any scaling analysis for the RRG model is severely complicated by the fact that the crossing point for finite-site scaling curves ($W$-dependencies of some observable for different fixed $L$) drifts strongly with $L$ \cite{tikhonov2016anderson}. In fact, a trend towards such a behavior has been also observed in the analysis of the Anderson transition $d$-dimensional systems with $d > 3$
\cite{zharekeshev1998kh,markovs2006numerical,garcia2007dimensional,ueoka2014dimensional,tarquini2017critical}; it becomes particularly pronounced for the largest studied value $d=6$. Another manifestation of large corrections to scaling is a strong variation of the flowing exponent $\nu_{\rm del}(\tau)$, see Fig.~\ref{fig:cvol}b.

Since the RRG model realizes, in a certain sense, a $d=\infty$ limit of the Anderson transition, one can wonder whether a ``symmetric'' scaling  (the same exponent $\nu$ on both sides) in a finite $d$ is not in conflict with an ``asymmetric'' scaling ($\nu_{\rm del} \ne \nu_{\rm loc}$) on RRG. The resolution is that the range of validity of the symmetric scaling around the critical disorder gradually shrinks towards zero with increasing dimensionality, $d\to \infty$.

As has been mentioned in Sec.~\ref{sec:introduction}, the RRG model can be viewed as a toy-model for the MBL transition. Let us briefly discuss the known numerical results for the scaling near the MBL transition  (which were mainly obtained for interacting 1D spin chains with disorder in the form of a random Zeeman field \cite{oganesyan07,kjall2014many,nandkishore2015many,luitz2015many,serbyn2015criterion,geraedts2017characterizing,khemani2017critical,abanin2017recent,doggen2018many,mace2018multifractal}) and compare them to the Anderson transition on RRG. The two types of transitions show indeed a great deal of similarity. The tentative position of the MBL transition as derived from the data for systems of size $L$ moves strongly towards larger disorder, in full similarity with RRG. The critical point of the MBL transition appears to show, in many respects, properties of the localized phase, also in similarity with RRG. Furthermore, the critical indices that have been found by numerical scaling analysis of the MBL transition are in the range $0.5-1$~\cite{kjall2014many,luitz2015many,mace2018multifractal}. This shows again a similarity with the RRG transition, for which the index on the localized side is $\nu_{\rm loc}=1$ and on the delocalized side $\nu_{\rm del}=1/2$, with finite-size effects in small systems leading to an apparent increase of the latter value, see Fig.~\ref{fig:cvol}b.  We note that the above values are
 in a strong conflict with Harris criterion $\nu \ge 2/d$, so that they cannot be the true asymptotic exponents for the MBL transition. The tentative resolution of this apparent contradiction is that the true asymptotic behavior shows up only in very large systems $L\gtrsim 500-5000$~\cite{chandran2015finite}. Thus, systems of moderate sizes (relevant to experiments) may exhibit the behavior akin to that in the RRG model. (See Ref.~\cite{tikhonov18}   for ``translation''  of the RRG critical behavior to that with respect to the spatial length $L$; for the dimensionality $d=1$ this does not modify the exponents.)

 What are further lessons that results of the present work teach us in connection with investigation of the scaling at MBL transitions? First, the exponents controlling the finite-size scaling in the RRG problem are different on both sides of the transition ($\nu_{\rm del}=1/2$ vs. $\nu_{\rm loc}=1$). It is quite likely that the same property is relevant to the MBL transition as well. On the other hand, most of the previous analysis of numerical data around the MBL transition \cite{kjall2014many,luitz2015many}  has assumed equal exponents on both sides.  Very recently, a scaling analysis with different exponents has been carried out \cite{mace2018multifractal}, with the results $\nu_{\rm del} \simeq 0.45$ and $\nu_{\rm loc} \simeq 0.76$ quite similar to the RRG values.  A more detailed study of this ``asymmetry'' of the critical behavior at the MBL transition would be of great interest. Second, taking into account corrections to scaling is of crucial importance for proper analysis of this class of localization transition. Third, the accuracy of the numerical study of the transition in the present work was greatly favored by the possibility to apply the population-dynamics approach. As has been already emphasized above, this has  allowed us to increase the maximal effective system size from 20 (characteristic for the exact-diagonalization study) to more than 60.  Key to this progress in the study of the RRG transition was the relation to the infinite-Bethe-lattice model arising in the framework of the field-theoretical analysis. It would be important to see whether any more realistic MBL model allows for a similar treatment.  Finally, our observation that $L \approx 60$ is sufficient to reach the
 asymptotic behavior with a very reasonable accuracy (at least for the RRG model) is encouraging from the point of view of experimental studies of the MBL problems
 (for a recent review see Ref. \cite{abanin2018ergodicity}). Indeed, the reported experimental realizations with cold-atom systems contained about 100 of atoms \cite{choi2016exploring,schreiber2015observation}.  Further, quantum simulators with $\simeq 50$ particles
 based on Rydberg states of trapped cold atoms \cite{bernien2017probing} and on trapped ions \cite{smith2016many}  have been reported; such or related systems can also serve for the investigation of the MBL transition.

Finally, one foresees that quantum computers with 50-100 qubits will be available in a not so far future \cite{kelly2019operating}.
One can thus expect important results on critical properties near the MBL transition (even if describing only some intermediate regime) provided by quantum computations and simulations in the next few years. It is worth mentioning that the study of statistical properties of states on the ergodic side of the MBL transition has much in common with the problem of sampling from complex distributions that is an indicator of quantum supremacy that is expected to be reached when the number of qubits exceeds 48 \cite{boixo2018characterizing}.
\section{Acknowledgments}
We are grateful to M. V. Feigel'man for useful discussions and A. Scardicchio for attracting our attention to preprint \cite{parisi2018}. This work is supported by the program 0033-2019-0002 by the Ministry of Science and Higher Education of Russia. KT acknowledges support by Alexander von Humboldt Foundation.

\bibliography{rrg}

\begin{thebibliography}{68}
\bibitem [{foo()}]{foot-range}%
  \BibitemOpen
  \href@noop {} {}\bibinfo {note} {{For $M$ we consider the range
  $2^{20}-2^{30}$, for $b$ the range $2^{-11}-2^{-9}$, and for $h$ the range
  $1-4$.}}\BibitemShut {Stop}%
\end{thebibliography}

%%%%%%%%%%%%%%%%%%%%%%%%%%%%%%%%%%%%%%
%%%%%%%%%%%%%%%%%%%%%%%%%%%%%%%%%%%%%%
%%%%%%%%%%%%%%%%%%%%%%%%%%%%%%%%%%%%%%
%%%%%%%%%%%%%%%%%%%%%%%%%%%%%%%%%%%%%%
%%%%%%%%%%%%%%%%%%%%%%%%%%%%%%%%%%%%%%
%%%%%%%%%%%%%%%%%%%%%%%%%%%%%%%%%%%%%%
%%%%%%%%%%%%%%%%%%%%%%%%%%%%%%%%%%%%%%
%%%%%%%%%%%%%%%%%%%%%%%%%%%%%%%%%%%%%%
%%%%%%%%%%%%%%%%%%%%%%%%%%%%%%%%%%%%%%
\newpage
\onecolumngrid

\begin{center}
{ \bf \Large Supplemental Material\\[0.2cm] to the article ``Critical behavior at the localization transition on random regular graphs''\\[0.2cm]  by K.S. Tikhonov and A.D. Mirlin}
\end{center}

\setcounter{figure}{0}
\makeatletter
\renewcommand{\thefigure}{S\@arabic\c@figure}
\makeatother

\vskip1cm

In this Supplemental Material, we provide some additional information to the numerical analysis performed in the main part of the paper.

%%%%%%%%%%%%%%%%%%%%%%%%%%%%
\begin{figure}[h]
    \mbox{\includegraphics[width=0.30\linewidth]{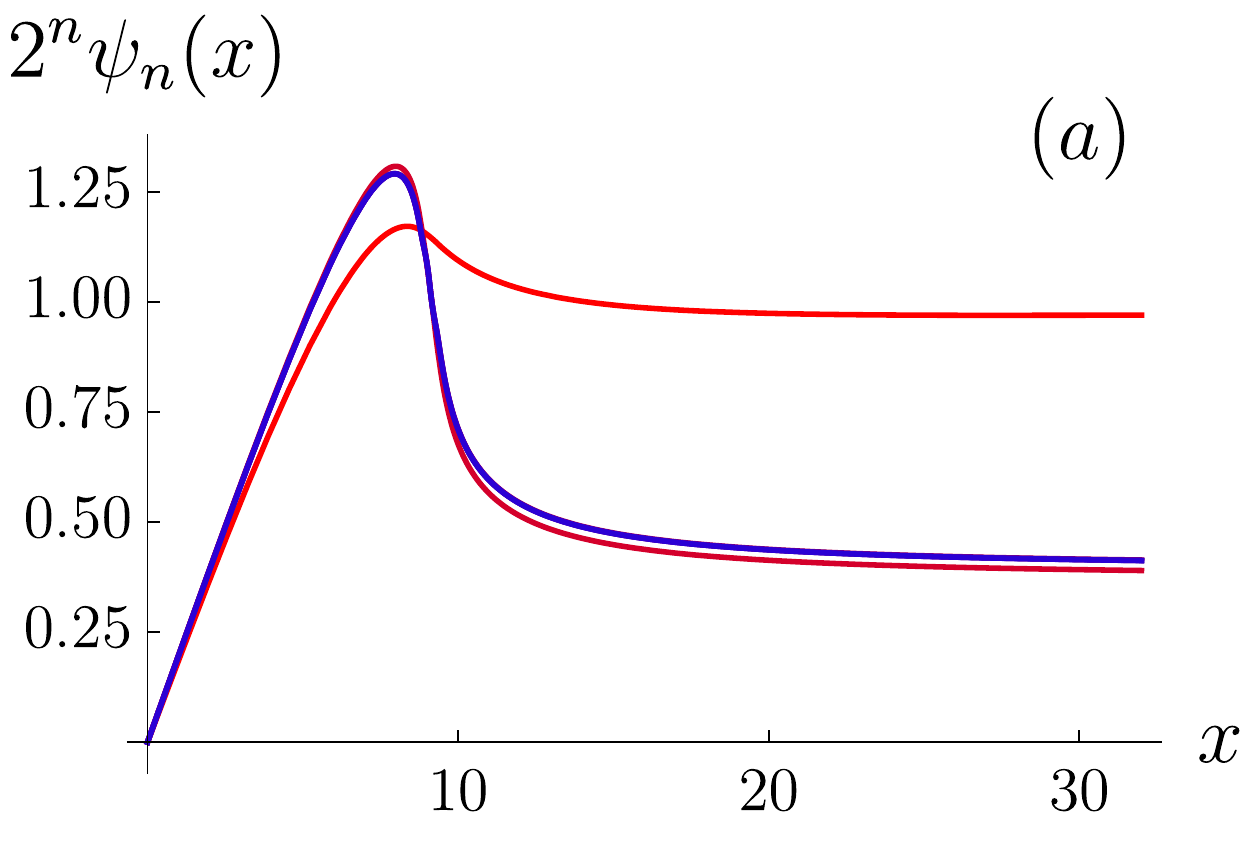}}
    \hspace{10px}
    \mbox{\includegraphics[width=0.30\linewidth]{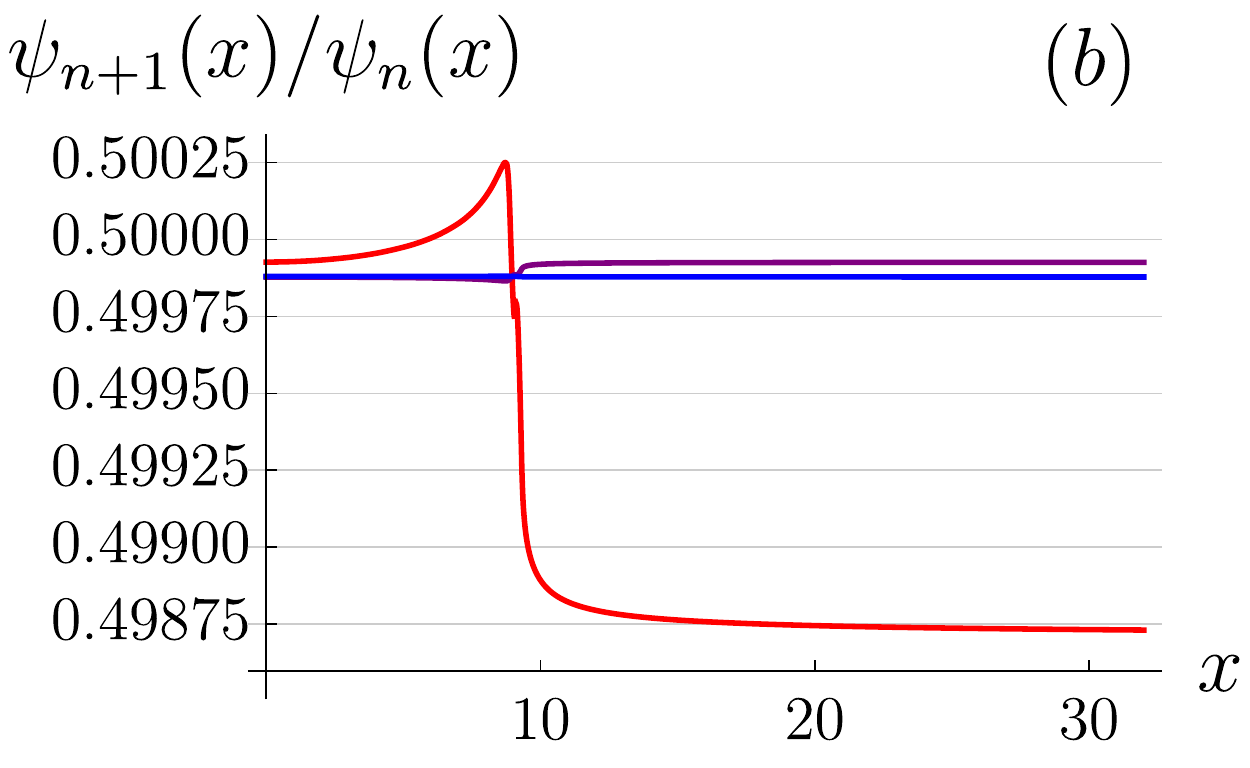}}
    \hspace{10px}
    \mbox{\includegraphics[width=0.30\linewidth]{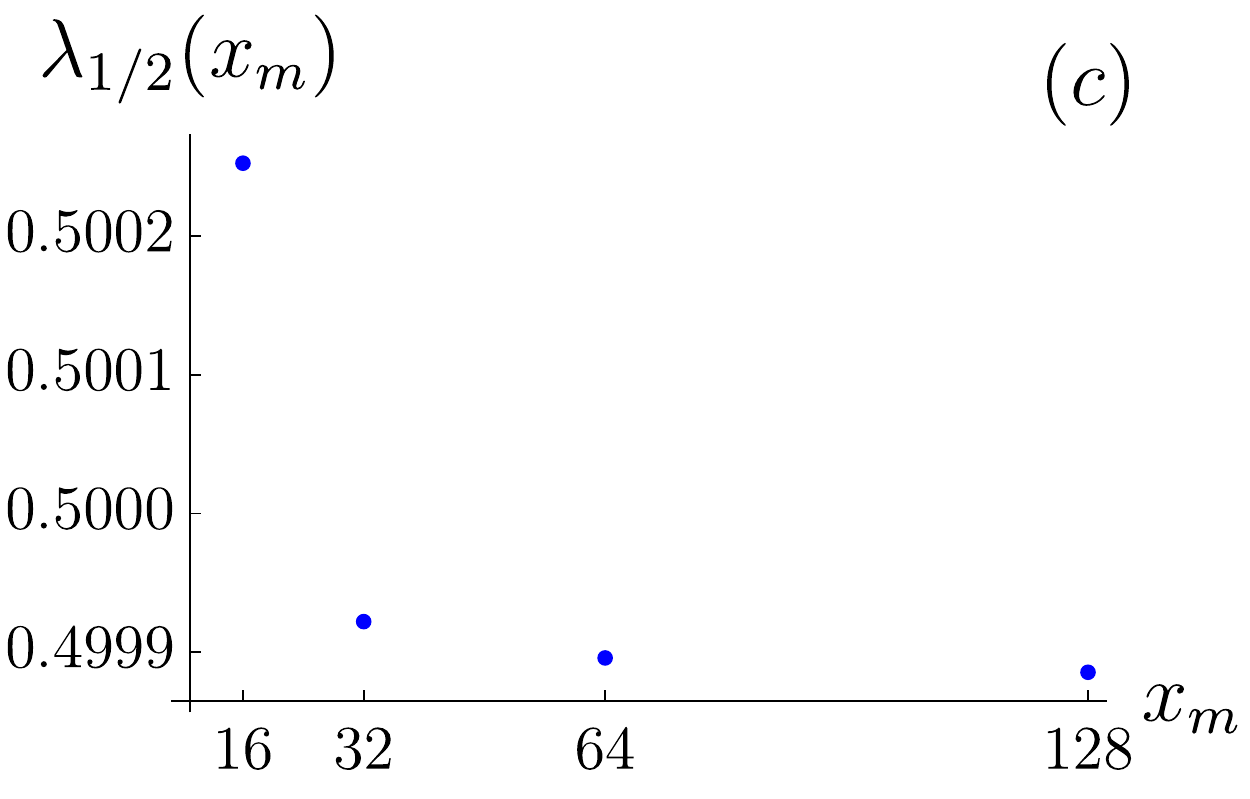}}
    \hspace{10px}
    \caption{Determination of the eigenvalue $\lambda_{1/2}$ for $W=18.175$.  (a)  First six iterations of Eq. (11) of the main text: $2^n\psi_n(x)$ for $n=0,\ldots,6$ (from red to blue). (b) Ratios $\psi_{n+1}(x)/\psi_n(x)$ for $n=3,4,5$ (from red to blue). (c) Dependence of the eigenvalue $\lambda_{1/2}$ on $x_m$.}
    \label{fig:it}
\end{figure}
%%%%%%%%%%%%%%%%%%%%%%%%%%%%

In Fig.~\ref{fig:it} we illustrate the analysis of numerical errors in course of computation of the critical disorder $W_c$, Sec. III of the paper. In particular, Fig.~\ref{fig:it}a shows results of the first six iterations of Eq. (11). The convergence is so fast, that they cannot be distinguished any more in this plot, starting from the fourth one. To demonstrate the accuracy quantitatively, we plot in Fig.~\ref{fig:it}b the ratio of two consecutive iterations $\psi_{n+1}(x)/\psi_n(x)$ starting from $n=3$. It is seen that after six iterations a convergence with 5-digit accuracy is reached. In Fig.~\ref{fig:it}c we show the convergence with respect to the parameter $x_m$ beyond which the asymptotic formula (12) is used. It is seen that $x_m=128$ (which is the value that we use in final calculations) provides a five-digit accuracy.

%%%%%%%%%%%%%%%%%%%%%%%%%%%%
\begin{figure}[tbp]
\includegraphics[width=\textwidth]{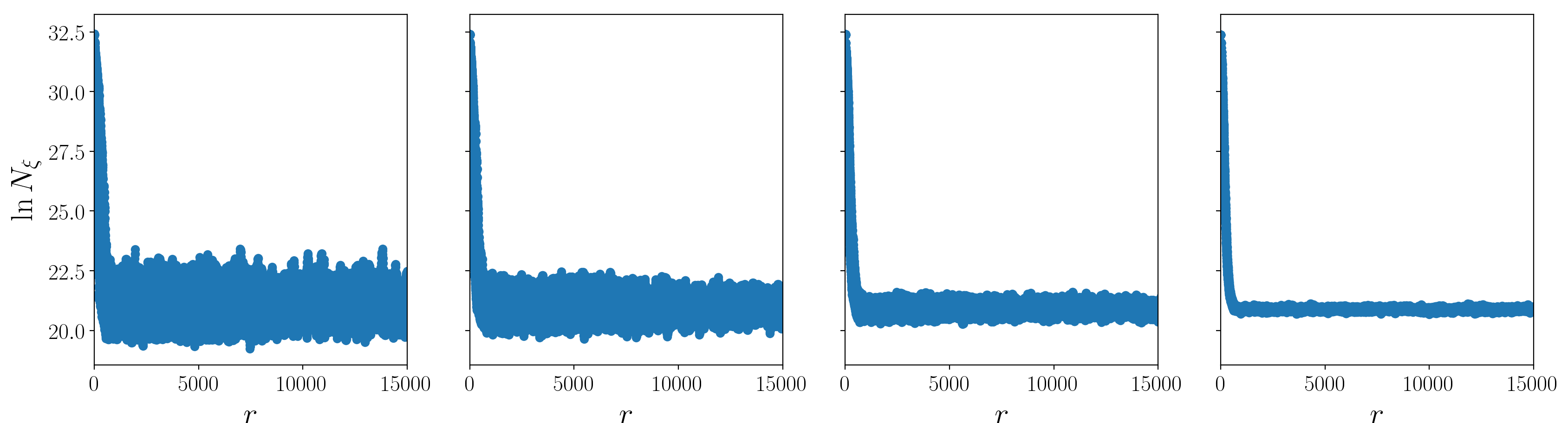}
\caption{Evolution of $N_\xi$ upon iteration of self-consistency equations at $W=16.5$ and various pool sizes $M=2^m$. Left to right: $m=18,19,20,24$ (several runs of the iterative procedure are combined).}
\label{fig:rdep}
\end{figure}
%%%%%%%%%%%%%%%%%%%%%%%%%%%%

Figures \ref{fig:rdep} and \ref{fig:mdepsup} provide a supporting information to the calculation of the correlation volume $N_\xi$ in Sec. IV of the main text.  In Fig.~\ref{fig:rdep} we illustrate the evolution of the corresponding population-dynamics iterative procedure for the disorder $W=16.5$ with the increasing pool size $M$ (from $2^{18}$ to $2^{24}$). For each value of $M$, 15000 iterations are shown.  A rather quick convergence with increasing $M$ is seen. This convergence is quantified in Fig.~\ref{fig:mdepsup}a where the resulting values of $\ln N_\xi (M)$ (averaged over fluctuations) are shown, with a fit to Eq. (15) of the main text. See also an analogous plot for $W=17$ in Fig. 2a of the main text.  In Fig.~\ref{fig:mdepsup}b we illustrate the dependence of $N_\xi (\eta, M)$  on the imaginary part of frequency $\eta$ (at fixed $M$). A convergence is quickly reached at $\eta \ll N_\xi^{-1}$.

%%%%%%%%%%%%%%%%%%%%%%%%%%%%
\begin{figure}[tbp]
\mbox{\includegraphics[width=0.45\textwidth]{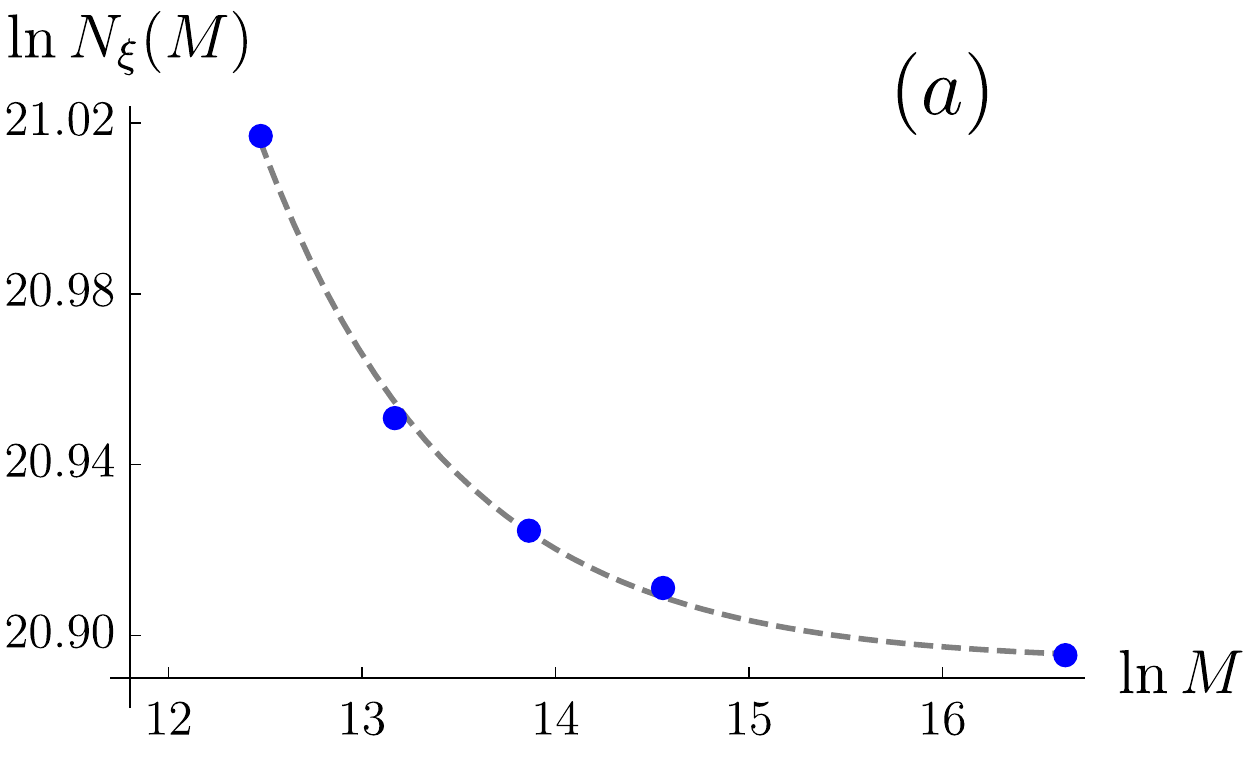}}
\hspace{10px}
\mbox{\includegraphics[width=0.45\textwidth]{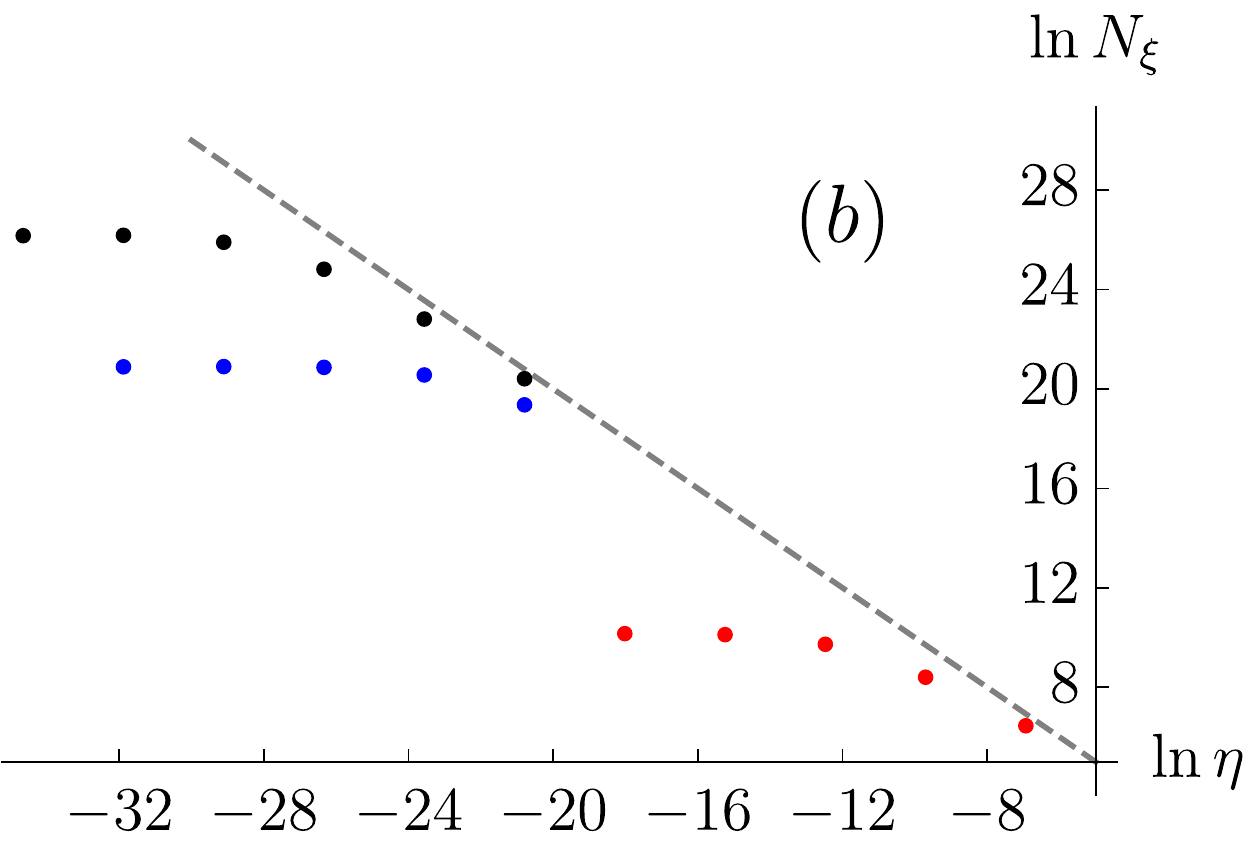}}
\caption{(a) Pool size-dependence of $N_{\xi}$ for $W=16.5$, see the data in Fig. \ref{fig:rdep} (compare also with Fig. 2a of the main text). The dashed line is a fit to Eq. (15) of the main text.  (b) $\eta$-dependence
of $N_\xi$ at fixed pool size $M=2^{28}$ for $W=17.0,\;16.5,\;14.0$ (black, blue, red), dashed grey line: $N_\xi=1/\eta$.}
\label{fig:mdepsup}
\end{figure}
%%%%%%%%%%%%%%%%%%%%%%%%%%%%

%%%%%%%%%%%%%%%%%%%%%%%%%%%%
\begin{figure}[tbp]
\mbox{\includegraphics[width=0.45\textwidth]{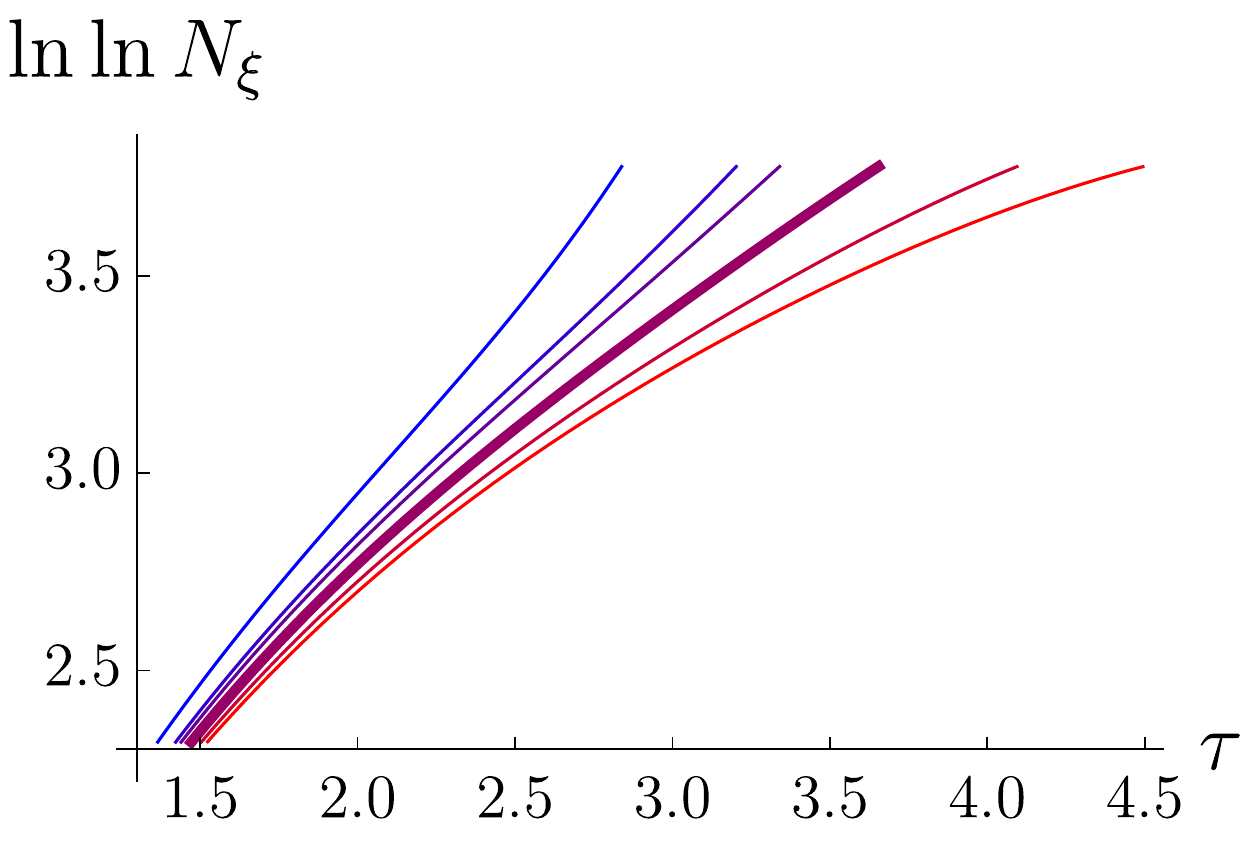}}
\hspace{10px}
\mbox{\includegraphics[width=0.45\textwidth]{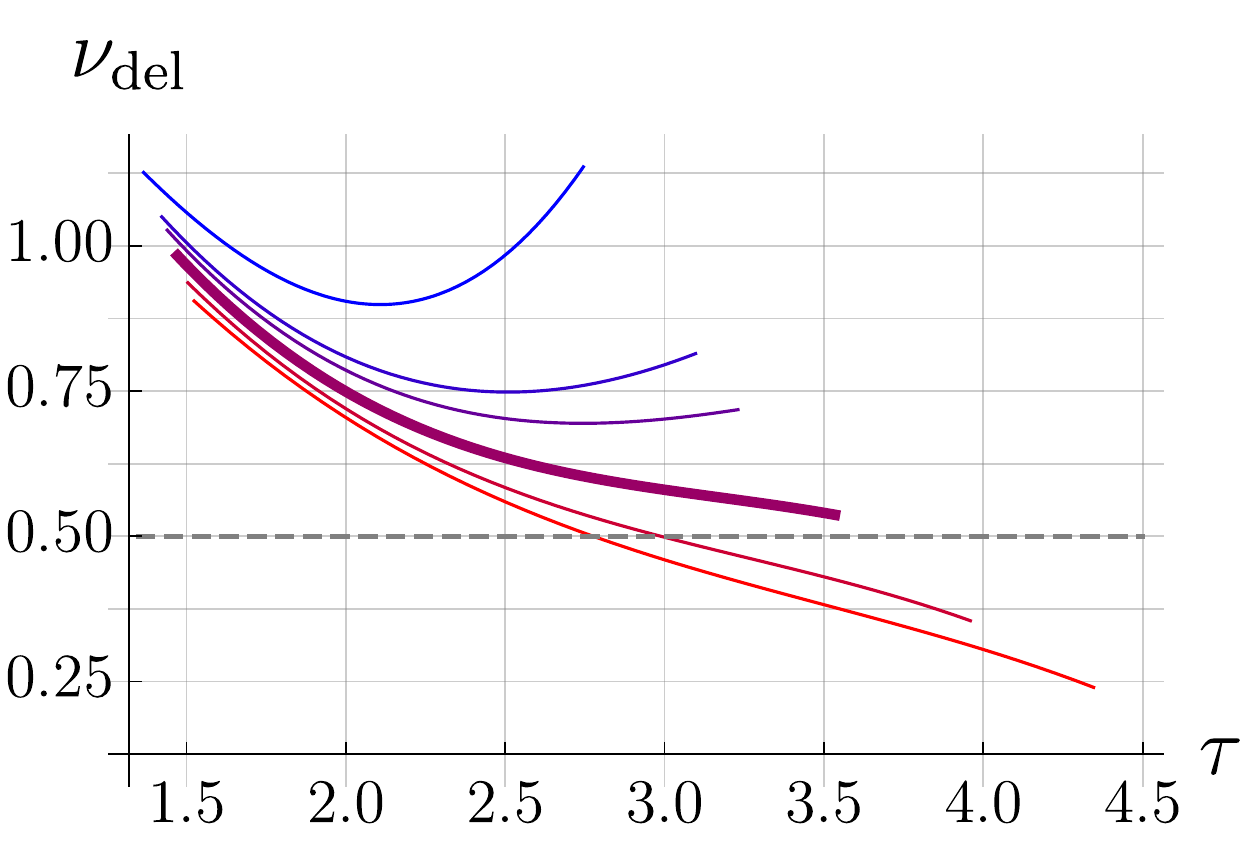}}
\caption{Plot analogous to Fig. 3 of the main text but with various values of the critical disorder $W_c$ used in the definition of the variable $\tau = -\ln(1-W/W_c)$.
From bottom to top (i.e., from red to blue): $W_c = 17.9,18.0,18.17,18.35,18.45, 18.8$. In the right panel (where the ``running critical exponent'' is shown), the thick line corresponding to the correct value $W_c=18.17$ shows a saturation towards the asymptotic value (true exponent) $\nu_{\rm del}=1/2$.  Lines with smaller $W_c$ approach zero, while those with larger $W_c$ tend to diverge.}
\label{fig:mod}
\end{figure}
%%%%%%%%%%%%%%%%%%%%%%%%%%%%

Figure \ref{fig:mod} illustrates the importance of the correct choice of the critical disorder $W_c$ in the definition of the scaling variable $\tau = -\ln(1-W/W_c)$ for the analysis of the critical behavior. It is analogous to Fig. 3 of the main text but with various values of the critical disorder $W_c$ used in the definition of the variable $\tau = -\ln(1-W/W_c)$. The right panel shows a derivative of curves in the left panel that yields the running critical exponent $\nu_{\rm del}(\tau)$. For the correct choice of the critical disorder $W_c=18.17$, this curve saturates to the asymptotic value $\nu_{\rm del}=1/2$. If a different  $W_c$ is used, the curve tends asymptotically to zero ($W_c$ smaller than the correct one such as 17.9 and 18.0 in the figure) or to infinity ($W_c$ larger than the correct one, such as 18.35, 18.45 and 18.8 in the figure).

Note that the value 18.8 is the one that was proposed as the critical disorder in Ref. 28 of the main text. A quick look at the right panel of Fig. S4 make it obvious that the ``running exponent'' $\nu_{\rm del} (\tau)$ with this choice of tentative $W_c$ does not show a saturation but rather has a clear tendency to diverge. However, the authors of Ref. 28 did not plot $\nu_{\rm del} (\tau)$ but rather limited themselves by plotting $\ln\ln N_\xi(\tau)$, as in the left panel of our Fig. S4. Of course, the variation of $\nu_{\rm del} (\tau)$ is contained in the curvature of $\ln\ln N_\xi(\tau)$ but is less evident in this representation. The authors of Ref. 28  ignored this and simply
fitted $\ln\ln N_\xi$ vs $\tau$ by a straight line in a wide range of disorder, $W=14$--$17.3$ and obtained a slope close to unity. The deviations from a straight line fit for the curve corresponding to $W_c=18.8$ in the left panel of Fig.~\ref{fig:mod} might
seem small in the corresponding range $\tau=1.4$--$2.5$. However, this apparent smallness is rather deceptive. Since this is a plot for $\ln \ln N_\xi$ variable, these deviations translate to more than 50\% in terms of the original variable $N_\xi$ (i.e., $\Im G$).
Clearly, this procedure is not particularly meaningful and can produce a rather arbitrary value of the slope, depending on the assumed value of $W_c$ and the chosen interval.
Additionally, as has been pointed out in Sec. III of the main text, the data for $N_\xi$ of Ref. 28 appear to be in addition plagued by numerical errors. One possible origin is that the authors of Ref. 28 used insufficiently small values of $\eta$, see the discussion in the end of Sec. III for more detail.

Finally, we provide a refinement to the Eq. (25) of the main text, including a subdominant term:
\be
\label{cvolthSM}
\frac{1}{\ln N_\xi}=0.0313\left(W_c-W\right)^{1/2}+c_{\textrm{fit}}\left(W_c-W\right)^{3/2},\;c_{\textrm{fit}}=0.00369.
\ee
The coefficient 0.0313 of the leading term was determined from the analytical theory supplemented by a numerical calculation of coefficients $c_1$ and $c_2$ characterizing the eigenvalue $\lambda_\beta$, see Eqs. (21), (22)  and (24) of the main text. The coefficient $c_{\textrm{fit}}$ of the subleading correction is determined by fitting to the data shown in Fig. 3a of the main text. This equation provides an excellent fit for the correlation volume in a range of $14$ orders of magnitude (see inset in Fig. 3a of the main text, blue dashed line).
It is worth emphasizing two points. First, the subleading correction is expected to depend on the way we define $N_\xi$. We recall that in this work it was numerically determined as $\exp\langle - \ln\Im G\rangle$.  Other possible definitions (based on moments of $\Im G$) will have the same leading behavior but the correction may vary. Second, a general expectation would be that a correction to the leading behavior of $\ln N_\xi$ has a relative smallness $\sim (W_c-W)^{1/2}$.  Interestingly, with $\ln N_\xi$ defined as $\langle - \ln \Im G\rangle$, we do not observe numerically such a correction (or, else, the corresponding coefficient is very small). The observed correction has a relative smallness $W_c-W$. 

\end{document}